\documentclass[%
preprint,
superscriptaddress,
groupedaddress,
nofootinbib,
 amsmath,amssymb,
 aip,
showkeys
]{revtex4-2}

\usepackage{capt-of, float}
\usepackage{graphicx, bm, bbm, amsbsy,multirow,microtype,siunitx}
\usepackage{array,xcolor}

\usepackage{hyperref}

\newcommand{\bq}{\begin{equation}}
\newcommand{\eq}{\end{equation}}
\newcommand{\bqs}{\begin{equation*}}
\newcommand{\eqs}{\end{equation*}}
\newcommand{\bqa}{\begin{eqnarray}}
\newcommand{\eqa}{\end{eqnarray}}
\newcommand{\bqas}{\begin{eqnarray*}}
\newcommand{\eqas}{\end{eqnarray*}}
\def\etal{{\em et al.\ }}

\newcommand{\vu}{{\mathbf{v}}}
\newcommand{\vn}{{\mathbf{n}}}
\newcommand{\vD}{{\mathbf{D}}}
\newcommand{\vDp}{{\mathbf{D'}}}
\newcommand{\Dfig}[2]{\includegraphics*[width=#2in]{#1}}
\newcommand{\Pfig}[2]{\includegraphics*[width=#2in]{#1.png}}

\newcommand{\Rey}{\textit{Re}}
\newcommand{\vx}{{\mathbf{x}}}
\newcommand{\va}{{\mathbf{a}}}
\newcommand{\vecr}{{\mathbf{r}}}
\newcommand{\vxi}{\boldsymbol{\xi}}
\newcommand{\vsigma}{\boldsymbol{\sigma}}
\newcommand{\vlambda}{\boldsymbol{\lambda}}

\newcommand{\vq}{{\mathbf{q}}}

\newcommand{\trans}{\mathsf{T}}

\newenvironment{Eoneplots}[2]
{
\begin{figure}[htpb!]
          \begin{center}
              \include{#1}
           \vspace{-.25in}
       \caption{ {\footnotesize #2}}
          \label{#1}
          \end{center}
        \vspace{-0in}
\end{figure}
}{}
\newenvironment{Eoneplotsw}[2]
{
\begin{figure}[htp]
          \begin{center}
              \include{#1}
           \vspace{-0in}
       \caption{ {\footnotesize #2}}
          \label{#1}
          \end{center}
        \vspace{-0in}
\end{figure}
}{}
\newenvironment{Etwofigs}[5]
{
        \begin{figure} [htpb]
          \begin{center}
          \setlength{\tabcolsep}{.1in}
          \begin{tabular}{cc}
          \multicolumn{1}{l}{$\bf{(a)}$}&\multicolumn{1}{l}{$\bf{(b)}$}\\
              \Dfig{#1}{#4} & \Dfig{#2}{#5} \\
                   \vspace{-.25in}
          \end{tabular}
          \caption{\footnotesize #3}
          \label{#1}\label{#2}
          \end{center}
        \vspace{-.25in}
        \end{figure}
}{}

\def\XXint#1#2#3{{\setbox0=\hbox{$#1{#2#3}{\int}$}
\vcenter{\hbox{$#2#3$}}\kern-.5\wd0}}

\DeclareMathOperator{\diag}{diag}
\DeclareMathOperator{\tr}{tr}
\pdfoutput=1

\begin{document}
\title{An incompressible Eulerian method for fluid-structure interaction with mixed soft and rigid solids}

\author{Xiaolin Wang}
\email{xiaolinwang@seas.harvard.edu}
\affiliation{John A. Paulson School of Engineering and Applied Sciences, Harvard University, Cambridge, MA 02138, USA}
\affiliation{Department of Mechanical Engineering, Massachusetts Institute of Technology, Cambridge, MA 02139, USA}
\author{Ken Kamrin}
\email{kkamrin@mit.edu}
\affiliation{Department of Mechanical Engineering, Massachusetts Institute of Technology, Cambridge, MA 02139, USA}
\author{Chris H. Rycroft}
\email{chr@seas.harvard.edu}
\affiliation{John A. Paulson School of Engineering and Applied Sciences, Harvard University, Cambridge, MA 02138, USA}
\affiliation{Mathematics Group, Lawrence Berkeley National Laboratory, 1 Cyclotron Road, Berkeley, CA 94720, USA}

\date{\today}

\begin{abstract}
  We present a general simulation approach for incompressible fluid--structure interactions in a fully Eulerian framework using the reference map technique (RMT). The approach is suitable for modeling one or more rigid or finitely-deformable objects or soft objects with rigid components interacting with the fluid and with each other. It is also extended to control the kinematics of structures in fluids. The model is based on our previous Eulerian fluid--soft solver \cite{rycroft2018reference}, and generalized to rigid structures by constraining the deformation-rate tensor in a projection framework. Several numerical examples are presented to illustrate the capability of the method.
\end{abstract}

\maketitle

\section{Introduction}
Understanding fluid--structure interaction (FSI) has always been an essential research topic to the computational fluids community due to its many applications in various disciplines. Since fluids and solids are often discretized with different grids, considerable attention has been paid to the accurate description of fluid--structure coupling. One set of FSI approaches treats the fluid on a fixed Eulerian mesh and the structure with Lagrangian points, such as the family of the immersed boundary methods~\cite{peskin2002immersed,mittal2005immersed}, which provides a framework for coupling fluids and rigid~\cite{uhlmann2005immersed, seo2011sharp, yang2009smoothing} or elastic bodies~\cite{fai2013immersed,griffith2009simulating, huang2007simulation} using a smoothed delta function formulation. Another set of approaches uses Lagrangian description for both fluids and structures, and uses an arbitrary Lagrangian--Eulerian (ALE) method to avoid excessive distortion of meshes. This approach has been successfully employed in fluids coupled to finite structural deformations and rigid objects~\cite{farhat1998load, hu2001direct,rugonyi2001finite}. Meshless formulations or particle-based methods including the smoothed-particle hydrodynamics technique (SPH)~\cite{antoci2007numerical,monaghan2005smoothed}, material point method (MPM)~\cite{lian2011coupling} and moving particle semi-implicit method (MPS)~\cite{you2016mps,sun2015modified} have also been investigated, particularly for cases with free fluid surfaces.

In recent years, a fully Eulerian-frame FSI approach has been developed called the reference map technique (RMT)~\cite{kamrin2012reference,valkov2015eulerian,rycroft2018reference,kamrin2008stochastic}, to describe nonlinear material deformation on fixed Eulerian grids. The following features have been implemented into RMT-based FSI simulation: (i) an Eulerian-frame solid formulation for constructing constitutive response of nonlinear material models using the reference map field to track deformation~\cite{kamrin2012reference, valkov2015eulerian}, (ii) a strong coupling relation, both blurred and sharp, that moves the fluid--structure interface simultaneously on fixed Eulerian grids, (iii) a discrete conservative formulation~\cite{kamrin2012reference,jain2019conservative}, (iv) a robust and spatially second-order accurate solver for incompressible fluids and solids~\cite{rycroft2018reference} using the projection method framework of Chorin~\cite{chorin1967numerical, chorin1968numerical}, (iv) advanced solid simulation features such as the ability to simulate sharp corners on finite grids using auxiliary level-set function, multiple object contacts, and actuation~\cite{rycroft2018reference}. There are still several aspects needed to complete the whole Eulerian-framework FSI solver. First, our Eulerian approach relies on the material deformation stress to determine the simultaneous motion of fluid and structure, and lacks a natural generalization to fluid--rigid interactions where the structure undergoes no deformation. This limits the use of our method for problems involving rigid structures. Although we can model rigid objects as stiff ones with very large elastic modulus, and take advantage of the existing fluid-soft solver, this requires very small time steps since we implemented the soft stress term explicitly, which makes it infeasible to use for long simulations. Therefore, we prefer to construct a direct fluid-rigid solver under the same Eulerian framework that allows a larger time step. Secondly, we currently control the kinematics of a structure, such as an actuation motion, through internal body stress, which could be difficult to compute in advance if we want to assign certain velocity or trajectory profiles to the structure. Finally, many research problems in material and biomechanical applications involve deformable structures with rigid components, which is difficult to simulate with most existing methods. Some examples include adaptive composite marine propulsors and turbines where deformable and passively-controlled composite are attached to rigid marine structures for better performance \cite{young2016adaptive}, structural dynamics of plant leaves where stems and leaves have significant distinct stiffness \cite{miller2005structural,speck2011plant}, and the motion of skeletal muscles and cartilage where deformable tissues are attached to rigid bones. In this paper, we present such a mixed fluid--soft and rigid body interaction model that extends the previous Eulerian framework and can easily incorporate different features mentioned above in a single simulation. We are also capable of controlling the kinematic condition of objects more accurately with this model.

In the existing literature, the simulation approach for fluid--rigid interactions can be put into two major categories: partitioned approaches, where the fluid and structure domains are solved separately and then coupled together~\cite{causin2005added, badia2008fluid,uhlmann2005immersed,kempe2012improved}, and monolithic approaches, where equations for the fluid and the rigid structure are solved simultaneously~\cite{robinson2011symmetric,gibou2012efficient,batty2007fast,apte2009numerical}.
The partitioned approach is typically formulated in a staggered fashion with fluid pressure imposed as a direct forcing on the structure and the structure velocity imposed as boundary conditions to solve the fluid equation. It has been employed under different frameworks, including the immersed boundary method~\cite{uhlmann2005immersed,kempe2012improved}, with the capability using the existing fluid and structure solvers. A disadvantage is that it is difficult to avoid the numerically spurious oscillation in the pressure field, although special treatments can reduce this effect on the results~\cite{seo2011sharp}. The monolitic coupling approaches have the potential to be more stable, but formulating and solving a discretized system of equations for the monolithic system is not an easy task.

In recent years, several efforts have focused on formulating the fluid--rigid coupling relation in special forms so the final linear system can be efficiently computed. Robinson-Mosher \etal designed a monolithic approach based on a projection framework, which leads to a symmetric positive definite (SPD) system in the projection step~\cite{robinson2011symmetric}. The method was applied to both rigid objects and elastic structures with a linear constitutive law. Gibou and Min designed a fractional approach where the interactions between fluids and solids are enforced via a projection step. Their scheme also produced a SPD linear system in the projection step, which can be solved efficiently using standard techniques such as preconditioned conjugate gradient method~\cite{gibou2012efficient}. Gr{\'e}tarsson \etal developed an implicit coupling system for Eulerian compressible fluid and volumetric Lagrangian solids~\cite{gretarsson2011numerically}. The coupled interactions were formulated into a symmetric indefinite system, and was made SPD under more assumptions that conserved the momentum and kinematic energy.

In this work, we present a monolithic incompressible fluid--rigid coupled method on a fully Eulerian grid using the projection framework. The computation of intermediate step solutions follows our previous fluid--soft interaction solver~\cite{rycroft2018reference}. The incompressibility and rigidity requirements for fluids and solids are enforced simultaneously in the projection step via pressure and an artificial rigidity stress field. Our method uses a stress constraint instead of a force field to enforce the rigid structure motion, which naturally preserves the numerical momentum in the computation. The method employs a finite-element approximation projection approach~\cite{almgren1996numerical}, and is formulated into an SPD system in the projection step for computational efficiency. The idea of enforcing rigid motion inside the object through certain constraints has been implemented in different frameworks. Glowinski \etal presented a distributed Lagrange-multiplier (DLM) method, and enforced the rigid motion by restricting the velocity field to satisfy a rigid form~\cite{glowinski2001fictitious}. Later, Patankar \etal presented a new DLM formulation by constraining the deformation-rate tensor, which is similar to the constraint used in our model~\cite{patankar2000new}. Coquerelle and Cottet provided a vorticty formulation to solve the problem and recovered the rigid motion inside an object via projection steps~\cite{coquerelle2008vortex}.

The paper has the following structure: Section \ref{sec:modeling} discusses the basic model in our work, with numerical details in Section \ref{sec:numeric}. In Section \ref{sec:results}, we present single rigid object examples and an example of elastic object with rigid components. We then present simulations with multiple rigid objects, with comparisons to other computational results, as well as mixed soft and rigid interactions. Finally, we discuss how to generalize the model to control the kinematics of structure motion in a fully coupled system via the projection framework.

\section{Modeling}
\label{sec:modeling}
\subsection{Overview of the fluid--structure interaction}
We consider solving a fully coupled fluid--structure interaction problem in a two dimensional incompressible fluid, as shown in Fig.~\ref{fig:schematic}(a). Here we first consider a simpler case where a rigid structure $\Omega_r$ and a soft deformable structure $\Omega_s$ are immersed in the fluid $\Omega_f$ without contact between two solids. The more general multiple body contact model is described later in Subsection \ref{sub:multi_c}. We call the whole computation domain $\Omega$.

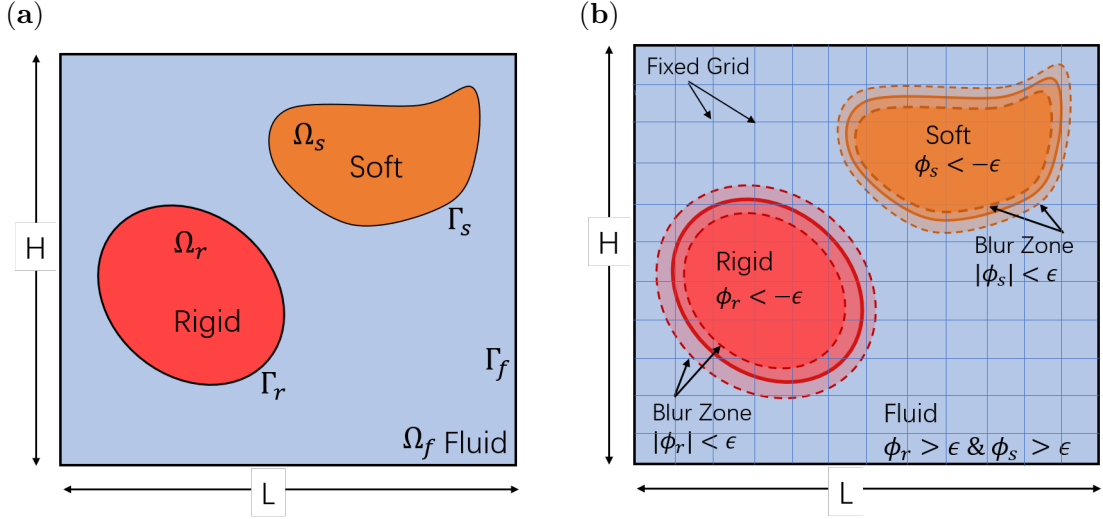
\Etwofigs{schematic}{blurzone}{\label{fig:schematic}(a) A schematic diagram of the fluid--structure interactions considered in this paper. $\Omega_f$ is the fluid domain and $\Gamma_f$ is the boundary of the computational domain. $\Omega_r$ is the rigid-solid domain with $\Gamma_r$ as the fluid--rigid interface. $\Omega_s$ is the soft-solid domain with $\Gamma_s$ as the fluid--soft interface. (b) Overview of the
reference map technique for simulating fluid--structure interaction on a fixed Eulerian grid. The zero contour of level set function $\phi_r(\vx,t)$ indicates the fluid--rigid interface. The zero contour of level set function $\phi_s(\vx,t)$ indicates the fluid--soft interface. The blur zone is defined as the region where $|\phi_r|<\epsilon$ and $|\phi_s|<\epsilon$, respectively, where $\epsilon$ is the blur width. It straddles the fluid--solid interface.}{2.8}{2.8}

In the computational domain, both fluid and structure satisfy a momentum balance equation,
\bq
\rho\left(\frac{\partial\vu}{\partial t}+\vu\cdot\nabla\vu\right)=-\nabla p+\nabla\cdot\vsigma,
\eq
and the incompressibility constraint,
\bq
\label{eq:i_constr}
\nabla\cdot\vu=0.
\eq
Here, $\rho(\vx,t)$ is the density, $\vu(\vx,t)$ is the velocity, $p(\vx,t)$ is the pressure, and $\vsigma(\vx,t)$ is the deviatoric part of the Cauchy stress tensor. All of these fields are defined globally and represented numerically on fixed Eulerian rectangular grids, but they take different expressions based on which domain the grid point $\vx$ lies in. The divergence of the stress tensor is defined as $(\nabla\cdot\vsigma)_i = \sum\limits_j\partial\sigma_{ij}/\partial x_j$ for $i,j=1,2$ in two dimensions.

For each object, we introduce a level set function $\phi(\vx,t)$ whose zero contour corresponds to the fluid--solid interface~\cite{sethian1996level, osher2004level}, with the convention that $\phi<0$ in the solid domain and $\phi>0$ in the fluid domain. A detailed discussion of evaluating $\phi(\vx,t)$ at each grid point is presented in Subsection \ref{sub:rmt}. Using the level set function, we can define a smoothed Heaviside function $H_\epsilon(\phi)$ with a transition region of width $2\epsilon$,
\bq
 H_\epsilon(\phi)=\left\{\begin{array}{ll}
     0 & \qquad\text{if $\phi\le-\epsilon$,}\\
     \frac{1}{2}(1+\frac{\phi}{\epsilon}+\frac{1}{\pi}\sin\frac{\pi\phi}{\epsilon})& \qquad \text{if $|\phi|<\epsilon$,}\\
     1 & \qquad \text{otherwise.}
 \end{array}\right.
\eq
This is a twice-differentiable form of the smoothed Heaviside function, which has been used in previous work~\cite{sussman94,sussman97,yu2003coupled}. The choice of $\epsilon$ and the form of $H_\epsilon(x)$ were discussed in detail by Rycroft \etal\cite{rycroft2018reference}. We define the region satisfying $-\epsilon<\phi<\epsilon$ as the \emph{blur zone}, which straddles the fluid--solid interface as shown in Fig.~\ref{fig:schematic}(b).

We now define level set fields $\phi_r$ for the rigid object and $\phi_s$ for the soft object respectively, which allows us to define a smooth density field among all phases by
\bq
\rho(\vx)=\rho_f(\vx)+(1-H_{\epsilon}(\phi_r))(\rho_r(\vx)-\rho_f(\vx))+(1-H_{\epsilon}(\phi_s))(\rho_s(\vx)-\rho_f(\vx)),\label{eq:density}
\eq
where $\rho_r(\vx)$, $\rho_s(\vx)$, and $\rho_f(\vx)$ are the density of rigid object, soft object and fluid respectively. In this work, we choose $\epsilon=2.5\Delta x$ for the examples, where $\Delta x$ is the grid spacing. Physically, we are interested in the limiting case where $\epsilon\rightarrow 0$ and a sharp interface between the fluid and solids is recovered. In our simulation, we choose $\epsilon$ to scale proportionally with the grid spacing, so that as the resolution of the method increases, the solution approaches this limit.

The deviatoric stress term $\vsigma$ takes different expressions in each domain. In the Newtonian fluid domain, we consider the viscous stress represented as $\vsigma_f = 2\mu_f \vD[\vu_f]$ where $\vD[\vu]=\frac{1}{2}(\nabla\vu +(\nabla\vu)^\trans)$ is the deformation rate tensor and $\mu_f$ is the kinematic viscosity of the fluid. For the soft deformable object, the deviatoric stress $\vsigma_s$ is constructed through constitutive relations of the material via the deformation gradient tensor. In particular, we consider the nonlinear neo-Hookean material for all the deformable examples in this work. The formulation of $\vsigma_s$ on a fixed Eulerian grid is discussed in Subsection \ref{sub:rmt}.

For rigid objects, we require the deformation rate of the object $\vD[\vu]$ to be zero throughout. This leads to an extra constraint that the velocity needs to satisfy in the rigid domain besides the incompressibility requirement:
\bq
\label{eq:r_constr}
\vDp[\vu]=\mathbf{0},
\eq
where we define $\vDp[\vu]=\vD[\vu]-\frac{1}{2}\tr(\vD[\vu])\mathbf{1} = \frac{1}{2}(\nabla\vu +(\nabla\vu)^\trans)-\frac{1}{2}(\nabla\cdot\vu)\mathbf{1}$ as the deviatoric part of the deformation rate tensor. Thus our numerical method must simultaneously satisfy the incompressibility constraint Eq.~\eqref{eq:i_constr} globally, and the rigidity constraint Eq.~\eqref{eq:r_constr} inside the rigid object.

We now interpret both incompressiblity and rigidity as extra constraints on the velocity field. The incompressibility constraint can be enforced when solving for the pressure, as discussed in Section \ref{sec:numeric}. Similarly, we assume there is a deviatoric rigid stress $\vsigma_r$ in the solid object to maintain the shear-free property of the solid. $\vsigma_r$ is symmetric and traceless, and can be written in the two-dimensional case as
\bq
\vsigma_r=\left(\begin{array}{cc}
\theta&\tau\\
\tau&-\theta
\end{array}\right)\, .
\label{eq:drigid}
\eq
We can similarly define a global deviatoric stress $\vsigma(\vx,t)$ on the computational domain that transitions between fluid and solid stresses,
\bq
 \vsigma(\vx)=\vsigma_f(\vx)+(1-H(\phi_r))(\vsigma_r(\vx)-\vsigma_f(\vx))+(1-H_{\epsilon}(\phi_s))(\vsigma_s(\vx)-\vsigma_f(\vx))\, . \label{eq:stress}
\eq
The soft-body stress $\vsigma_s$ and fluid stress $\vsigma_f$ are blurred over the blur zone. However, since we aim to enforce rigidity throughout the rigid solid, we make use of a true Heaviside function for the rigid stress term $\vsigma_r$. Therefore, $H(\phi_r)$ is the true Heaviside function and equals 0 at all grid points in the rigid domain, $\phi_r<0$. 

Altogether, the motion is updated in all phases by solving the following equations and constraints:
\begin{align}
\rho\left(\frac{\partial\vu}{\partial t}+\vu\cdot\nabla\vu\right)&=-\nabla p+\nabla\cdot\vsigma &&\text{for $\vx\in \Omega$,}\\
\nabla\cdot\vu&=0 &&\text{for $\vx\in \Omega$,}\\
\vDp[\vu]&=\mathbf{0} &&\text{for $\vx\in\Omega_r$.}
\end{align}

\subsection{Reference map technique}\label{sub:rmt}
To track the location and orientation of the moving object on a fixed grid, we introduce the reference map field $\vxi(\vx,t)$. At a current position $\vx$ and time $t$, the reference map $\vxi$ is defined as the initial position of the material that is now located at $\vx$. If the object is initially undeformed, then the field is initialized as $\vxi(\vx,0)=\vx$. The reference map field was originally employed in solid mechanics~\cite{gurtin2010mechanics,fachinotti2008finite,levin2011eulerian}, and recently has been generalized for solving soft deformable fluid--structure interactions within a fixed Eulerian grid framework~\cite{cottet2008eulerian,kamrin2008stochastic,valkov2015eulerian, kamrin2012reference,rycroft2018reference}. In particular, $\vxi(\vx,t)$ can be evolved with an advection equation,
\bq
\displaystyle\frac{\partial\vxi}{\partial t}+\vu\cdot\nabla\vxi= \mathbf{0}. {\label{eq:refmap}}
\eq
The two-dimensional deformation gradient tensor can be computed from the reference map field by ${\mathbf{F}}=(\nabla\vxi)^{-1}$ since the reference map field $\vxi$ is the inverse mapping of the motion function used in general continuum mechanics. Then the deformation gradient tensor can be used for further evaluation of structural stress through constitutive relations of the material. Though any hyperelastic model would follow the same procedure, in this work we consider the neo-Hookean solid model which, assuming plane-strain conditions, gives the in-plane stress as
\bq\label{NH}
\vsigma_s=G\, \left(\mathbf{B}-\frac{1}{3}\mathbf{1}(\tr\mathbf{(B)}+1)\right)
\eq
where $\mathbf{B}=\mathbf{F}\mathbf{F}^\trans$ represents the in-plane part of the left Cauchy--Green tensor\cite{rycroft2018reference}.

We note that the reference map field of a rigid body is solely determined by its center of mass motion $\vu_c(t)=(u_c(t),v_c(t))^\trans$ and angular velocity $\omega(t)$, and can be computed explicitly as
\bq
\vxi(\vx,t)
=\vx_c(0)
+{\mathbf{R}}(\Theta(t))^\trans \,  (\vx-\vx_c(t))\label{eq:updateref}
\eq
where $\vx_c(0)$ is the center of mass location in the initial configuration and $\vx_c(t)$ is the center of mass location in the current configuration which is computed as $\vx_c(t)=\vx_c(0) +\int_0^t \vu_c(t) dt$. In addition
\bq
\mathbf{R}(\Theta)=\left(\begin{array}{cc}
 \cos\Theta &-\sin\Theta\\
 \sin\Theta &\cos\Theta
 \end{array}
 \right)
\eq
rotates by $\Theta$, where $\Theta(t) = \Theta(0)+\int_0^t \omega(t)dt$ is the total angle through which the object has been rotated.

The level set field $\phi(\vx,t)$ can now be evaluated from the reference map field. In typical usage of level set methods~\cite{sethian1996level,osher2004level}, the level set field is time-integrated according to a hyperbolic partial differential equation, but here we use an alternative approach described in previous work~\cite{rycroft2018reference}. We first introduce a continuous function $\phi_0(\vxi)$ whose zero contour is the boundary of the object at the initial configuration, and $\phi<0$ indicates the solid structure. To update the level set field at each timestep, we first set the values so that $\phi(\vx,t)=\phi_0(\vxi(\vx,t))$. We then recover the signed distance property $|\nabla \phi|=1$ using a reinitialization procedure described by Rycroft and Gibou~\cite{rycroft2012simulations}. As shown in previous work~\cite{rycroft2018reference}, this procedure is reliable on complex geometries including those with sharp corners.

\section{Numerical Methods}\label{sec:numeric}
The numerical procedure introduces the rigidity constraint within the previous FSI framework of Rycroft \etal\cite{rycroft2018reference} for fluids and soft solids. The simulation domain is divided into an $M\times N$ grid
of rectangular cells of size $\Delta x$ by $\Delta y$. Following the work of Colella~\cite{colella1990multidimensional}, the velocity $\vu$, the reference map $\vxi$, and the level set $\phi$ are held at cell centers. Pressures $p$ and rigid stress components $\theta$ and $\tau$ are held at the cell corners, as shown in Fig.~\ref{fig:cellfigure}(a). In addition, the grid is padded by two
layers of cells in each direction whose values are populated to enforce different boundary conditions.

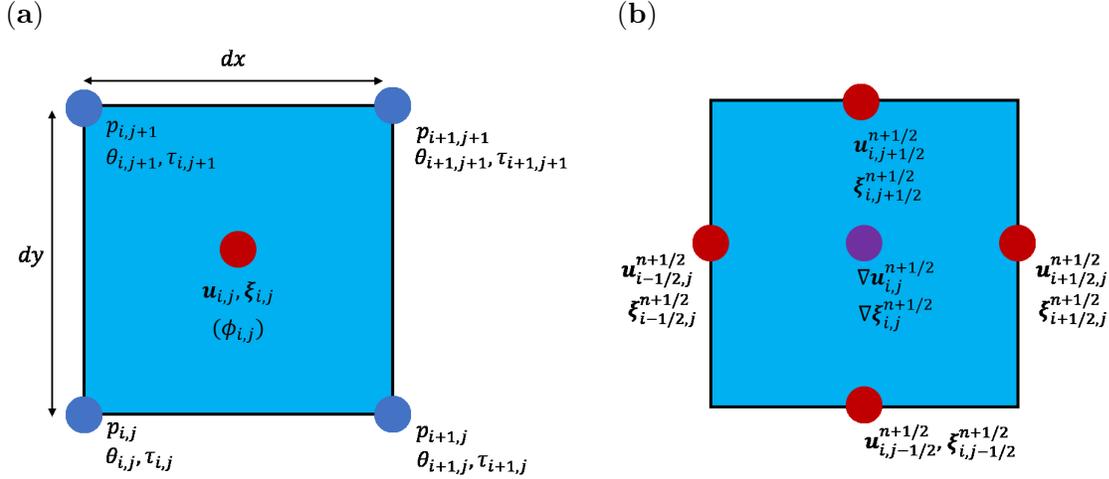
\Etwofigs{cellfigure1}{cellfigure2}{\label{fig:cellfigure}(a). Arrangement of the fields within a simulation grid cell. The reference map $\vxi_{i,j}$, velocity $\vu_{i,j}$ and the level set field $\phi_{i,j}$ are held at the cell center, while the pressure $p$ and the rigid stress components $\theta$ and $\tau$ are held at the cell corners. The level set field is derived from the reference map $\vxi$. (b). Arrangement of the edge velocities and reference maps that are computed at the half-timestep $n+1/2$. The resulting gradients are computed at the cell center using the central-difference scheme from edge variables. }{3.0}{3.0}

To advance the solution at every time step, we separate the momentum equations into two parts: an intermediate step velocity field $\vu^*$ that is computed through the explicit formula
\bq
\displaystyle\frac{\vu^*-\vu^n}{\Delta t} = \left(-\vu\cdot\nabla\vu\right)^{n+1/2}+\frac{1}{\rho (\phi^{n+\frac{1}{2}})}\nabla\cdot\vsigma^n,\label{eq:intereq}
\eq
and the pressure and rigid stress terms used to constrain the velocity field:
\bq
\displaystyle\frac{\vu^{n+1}-\vu^*}{\Delta t} = \frac{1}{\rho (\phi^{n+\frac{1}{2}})}\left(-\nabla p^{n+1}+\nabla\cdot\vsigma_r^{n+1}\right)\label{eq:proj-step2}.
\eq
The intermediate step solution $\vu^*$ is computed using the previous step's solutions, where the advective term $(\vu\cdot\nabla\vu)^{n+1/2}$ is evaluated at the middle of the time step using a second-order upwinding Godunov scheme, described in Subsection \ref{sub:adv}. In order to evaluate $\phi^{n+1/2}$ at the middle of the step, we also need to first evaluate $\vxi^{n+1/2}$, and then use the procedure described above to obtain $\phi^{n+1/2}$. We first obtain $\vxi^{n+1}$ by discretizing Eq.~\eqref{eq:refmap} as
\bq
\displaystyle\frac{\vxi^{n+1}-\vxi^{n}}{\Delta t}=-(\vu\cdot\nabla\vxi)^{n+1/2}\label{eq:interxi}
\eq
where the advective term $(\vu\cdot\nabla\vxi)^{n+1/2}$ is evaluated similarly using a second-order explicit Godunov scheme. We then obtain $\vxi^{n+\frac{1}{2}}=\frac{1}{2}(\vxi^n+\vxi^{n+1})$. The half-step reference map values $\vxi^{n+1/2}$ are updated with the same approach for both rigid and soft objects. For $\vxi^{n+1}$, Eq.~\eqref{eq:updateref} is instead used after the projection step to update the reference map value for rigid objects. We implemented this by first using a bicubic interpolation to obtain the velocity at the center of mass $\vu_c(t)$, and then obtain $\vx_c(t)$ with a second-order improved Euler method to compute the time integration.

Once the advective derivatives are evaluated, the intermediate velocity $\vu^*$ is computed using Eq.~\eqref{eq:intereq}. In order to evaluate the divergence of the stress deviator $\nabla\cdot\vsigma$ in Eq.~\eqref{eq:intereq}, the stress terms are first computed on the edge of each grid cell. The divergence is then computed as 
\bq
(\nabla\cdot\vsigma)_{i,j}=\displaystyle\frac{[\vsigma_x]_{i+1/2,j}-[\vsigma_x]_{i-1/2,j}}{\Delta x}+\frac{[\vsigma_y]_{i,j+1/2}-[\vsigma_y]_{i,j-1/2}}{\Delta y}
\eq 
where $[\vsigma_x]=(\sigma_{xx},\sigma_{xy})$ and $[\vsigma_y]=(\sigma_{xy},\sigma_{yy})$ are the components of the stress term acting on the vertical and horizontal edges, respectively.
In particular, the fluid stress deviator $\vsigma_f^n$ is computed using the previous solution $\vu^n$ and the soft solid stress deviator $\vsigma_s^{n+1/2}$ is evaluated using the neo-Hookean model (Eq.~\ref{NH}) where the deformation gradient is computed using $\vxi^{n+1/2}$. For more details of evaluating $\sigma_f$ and $\sigma_s$ on vertical and horizontal edges, we refer the readers in our previous work \cite{rycroft2018reference}. The rigid stress $\vsigma_r$ is computed at the projection step, which is not evaluated when computing $\vu^*$. Therefore, we take $\vsigma_r^n=\mathbf{0}$ when constructing the stress term $\vsigma^n$ in Eq.~\eqref{eq:intereq}. From here, we apply the incompressibility constraint ($\nabla\cdot\vu^{n+1}=0$) in the computational domain and the rigidity constraint ($\vDp[\vu^{n+1}]=\mathbf{0}$) in the rigid domain in Eq. ~\eqref{eq:proj-step2}. These lead to the following projection step equations:
\begin{align}
  \nabla\cdot\left(\frac{\Delta t}{\rho(\phi^{n+\frac{1}{2}})}\left(-\nabla p+\left[\begin{array}{c}\theta_x\\-\theta_y\end{array}\right]+\left[\begin{array}{c}\tau_y\\ \tau_x\end{array}\right]\right)^{n+1}\right) &= -\nabla\cdot\vu^* & \text{for $x\in\Omega$,} \label{eq:incomp}\\
  \vDp\left[\frac{\Delta t}{\rho(\phi^{n+\frac{1}{2}})}\left(-\nabla p^{n+1} +\nabla\cdot\vsigma_r^{n+1}\right)\right]&=-\vDp[\vu^*] & \text{for $x\in\Omega_r$.}\label{eq:rsd}
\end{align}
Equation \eqref{eq:rsd} can be further written as two independent equations,
\begin{align}
\nabla\cdot\left( \frac{\Delta t}{\rho(\phi^{n+\frac{1}{2}})}\left(\left[\begin{array}{c}p_x\\-p_y\end{array}\right]-\nabla\theta\right)^{n+1}\right) &= (u^*_x-v^*_y),\label{eq:dev1}\\
\nabla\cdot\left( \frac{\Delta t}{\rho(\phi^{n+\frac{1}{2}})}\left(\left[\begin{array}{c}p_y\\p_x\end{array}\right]-\nabla\tau\right)^{n+1}\right) &= (u^*_y+v^*_x).\label{eq:dev2}
\end{align}
In Eq.~\eqref{eq:incomp}, the rigid stress components $\theta$ and $\tau$ are defined to be zero outside the rigid domain, which is used as boundary conditions for Eqs.~\eqref{eq:dev1} \& \eqref{eq:dev2}. Outside the rigid domain, the first order derivatives of $\theta$ and $\tau$ are also zero, therefore Eq.~\eqref{eq:incomp} becomes the regular incompressibility equation commonly used in the fluid projection method, i.e.
\bq
-\nabla\cdot\left(\frac{\Delta t}{\rho(\phi^{n+\frac{1}{2}})}\nabla p\right) = -\nabla\cdot\vu^*.\label{eq:flincom}
\eq
Equations \eqref{eq:incomp}, \eqref{eq:dev1}, \& \eqref{eq:dev2} are solved using a finite-element formulation, described in Subsection \ref{sub:fep}. After this, the velocity is projected to be incompressible in the whole computational domain $\Omega$ and shear-free in the rigid domain $\Omega_r$ using
\bq
\vu^{n+1}=\vu^*+\displaystyle\frac{\Delta t}{\rho(\phi^{n+\frac{1}{2}})}\left(-\nabla p^{n+1}+\nabla\cdot\vsigma_r^{n+1}\right) \label{eq:updatev}
\eq
 where the gradient of $p^{n+1}$ and the divergence of $\vsigma_r^{n+1}$ are evaluated using a second-order centered difference formula at the cell center. We note that in the absence of rigid structures, this fluid-structure solver becomes the previous fluid--soft solver~\cite{rycroft2018reference} where Eq.~\eqref{eq:flincom} is solved at the projection step. 
\subsection{Advective terms} \label{sub:adv}
To evaluate the advective terms $(\vu\cdot\nabla\vu)^{n+1/2}$ and $(\vu\cdot\nabla\vxi)^{n+1/2}$ in Eqs.~\eqref{eq:intereq} \& \eqref{eq:interxi}, a second-order explicit Godunov scheme is used. This scheme was used in previous work~\cite{rycroft2018reference}, which contains a complete, detailed discussion.

When computing the advective terms, the cell centered velocities $\vu^n$ and reference maps $\vxi^n$ are first extrapolated to four cell edges at the mid-timestep $n+1/2$ using Taylor expansions, which are indexed using half-integers as shown in Fig.~\ref{fig:cellfigure}(b). After this step, each edge has velocities and reference maps from the two cells adjacent to it, and a Godunov upwinding procedure is used to select which values to use. We then perform an extra marker-and-cell (MAC) projection step on the edge velocities to ensure the incompressiblity of the edge velocities so that the discrete flux entering any grid cell is exactly zero~\cite{sussman99,yu2003coupled,rycroft2018reference}. The half-time gradients $\nabla\vu^{n+1/2}$ and $\nabla\vxi^{n+1/2}$ are then computed at the cell center using centered differences of the edge-based fields, after which the advective terms for the velocity and reference maps are evaluated as
\bq
(\vu\cdot\nabla a)^{n+1/2}_{i,j}=\displaystyle\frac{1}{2}\left(\begin{array}{c}u_{i-1/2,j}^{n+1/2}+u_{i+1/2,j}^{n+1/2}\\
v_{i,j-1/2}^{n+1/2}+v_{i,j+1/2}^{n+1/2}
\end{array}\right)\cdot\nabla a_{i,j}^{n+1/2}
\eq
where $a$ is a generic field component.
\subsection{Reference map extrapolation} \label{section:levelset}
The simulation makes use of a cell-centered level set function $\phi_{i,j}$ for tracking
fluid--solid interfaces, which is continually updated from the reference map field using the
procedure described in Section \ref{sub:rmt}. Computing the deviatoric stress term $\vsigma$ using Eq.~$\eqref{eq:stress}$ also requires computing the soft structure stress $\vsigma_s$ in the region $0<\phi_s<\epsilon$, which is beyond the true soft solid domain. However, the reference map field $\vxi$ only exists in the solid domain $\phi_s<0$. Therefore, we smoothly extend $\vxi$ into the region $0<\phi_s<\epsilon$ using extrapolation methods that will be described next. The value of the rigid stress $\vsigma_r$ does not use the reference map field. It is instead computed at the projection step. We still do such extrapolations to $\phi_r$ because we use a smooth density field as defined in Eq.~\eqref{eq:density} that requires level set values in $0<\phi_r<\epsilon$ beyond the rigid domain.

The extrapolation procedure works as follows~\cite{rycroft2018reference}. The grid points outside the solid domain are sorted and considered in increasing order of $\phi_s$. At each gridpoint a linear map for the reference map is constructed using least-squares regression, using all available existing $\vxi$ values in a $5\times 5$ box centered on the current gridpoint. After this, the reference map value at the current gridpoint is set by evaluating the linear map there. In rare cases, there may not be enough $\vxi$ values available to uniquely fit the linear map, in which case
the box is extended on all sides by one gridpoint and the procedure is attempted again. This procedure has proven effective and robust: it damps out high-frequency modes that could be the source of instability near the interface, and can successfully work even with deformable objects with sharp corners~\cite{rycroft2018reference}. For rigid objects, since we always update $\vxi^{n+1}$ using Eq.~\eqref{eq:updateref} at the end of the projection step, this extrapolation procedure still works well.

\subsection{Finite-element projection}\label{sub:fep}
 We use a finite-element formulation to solve Eqs.~\eqref{eq:incomp}, \eqref{eq:dev1}, \& \eqref{eq:dev2}. We only consider the case where the solids are fully immersed in the fluid domain. The pressure $p$ and the rigid stress components $\theta$ and $\tau$ are comprised of piecewise bilinear elements, and the velocity and density are
piecewise constant on the grid cells. For a given pressure element $\psi$ and rigid stress element $\gamma$, the weak formulation of Eqs.~\eqref{eq:incomp}, \eqref{eq:dev1}, \& \eqref{eq:dev2} is
\bqa
-\int_\Omega\frac{\Delta t}{\rho(\phi^{n+\frac{1}{2}})}\nabla p^{n+1}\cdot\nabla\psi \,dx\,dy+\int_{\Omega'_r}\frac{\Delta t}{\rho(\phi^{n+\frac{1}{2}})}\left(\left[\begin{array}{c}\theta_x\\-\theta_y\end{array}\right]+\left[\begin{array}{c}\tau_y\\ \tau_x\end{array}\right]\right)^{n+1}\cdot\nabla\psi\ dx\,dy \nonumber\\
= \int_\Omega-(\nabla\cdot\vu^*)\psi\ dx\,dy,\label{eq:weak1}\\
\int_{\Omega'_r}\frac{\Delta t}{\rho(\phi^{n+\frac{1}{2}})}\left(\left[\begin{array}{c}p_x\\-p_y\end{array}\right]-\nabla\theta\right)^{n+1} \cdot\nabla\gamma\ dx\,dy= \int_{\Omega'_r}(u^*_x-v^*_y)\gamma\ dx\,dy,\label{eq:weak2}\\
\int_{\Omega'_r}\frac{\Delta t}{\rho(\phi^{n+\frac{1}{2}})}\left(\left[\begin{array}{c}p_y\\p_x\end{array}\right]-\nabla\tau\right)^{n+1}\cdot\nabla\gamma\ dx\,dy = \int_{\Omega'_r}(u^*_y+v^*_x)\gamma\ dx\,dy.\label{eq:weak3}
\eqa
Here $\Omega'_r$ is the discretized analog of $\Omega_r$, and comprises of all of the grid cells that overlap some part of $\Omega_r$. Thus $\Omega_r \subseteq \Omega'_r$. We require $\gamma=0$ on $\partial\Omega'_r$, as the rigid stress vanishes at the fluid--solid interface. The resulting linear system is SPD (see Appendix \ref{section:spd} for detailed proof), and is solved using a preconditioned MINRES-QLP method~\cite{choi2011minres}. Although the original method was developed for an indefinite or singular matrix, we found it provided a more robust result for our system compared to other SPD Krylov solvers like conjugate gradient (CG) or SYMMLQ~\cite{paige75}. We also adopted the same strategy as suggested by Choi \etal \cite{choi2011minres} that when the condition number of the linear system is small (\textit{i.e.}~less than $10^6$), the MINRES scheme~\cite{paige75} is used to improve the computational efficiency.

A diagonal left-preconditioner matrix is implemented as
$M^{-1}=\diag\{L_\Omega^{-1},L_{\Omega'_r,0}^{-1},L_{\Omega'_r,0}^{-1}\}$. $L_\Omega$ is a Poission operator that can be efficiently solved with a multigrid method~\cite{Briggs2000,demmel1997applied}. As a preconditioner, we only require several fixed V-cycle iterations to obtain an approximate inverse matrix $\tilde{L}_\Omega^{-1}$. In this work, we use three V-cycles for the preconditioner. In order to preserve the SPD property of the original matrix, we modified the standard V-cycle to a symmetric version, where Gauss--Seidel sweeps with forward direction are performed in the downward part of the V-cycle, and sweeps with backward direction are performed in the upward part of the V-cycle. We implemented this by modifying the multithreaded custom C++ geometric multigrid library~\cite{rycroft2018reference,tgmg_website}. $L_{\Omega'_r,0}$ is a Poisson operator but on an irregular domain or disconnected domains if there are multiple rigid objects. Instead of a geometric multigrid solver, we use a CG solver and obtain an approximate $\tilde{L}_{\Omega'_r,0}$ as the preconditioner when the norm of the residual vector reaches a required tolerance $T_\text{CG}=10^6T\epsilon_m$ where $\epsilon_m$ is the machine epsilon for double precision floating point arithmetic and $T$ is the number of the rigid points.
\subsection{Multi-body contact} \label{sub:multi_c}
In previous work~\cite{rycroft2018reference}, we have illustrated the capability of the reference map technique to handle contacts between multiple objects. In summary, for $N$ soft objects, an independent reference map $\vxi^{(1)},\vxi^{(2)},\ldots, \vxi^{(N)}$ is introduced for each object. When the blur zones of two objects overlap, a stress addition within a small patch of the overlap zone is activated to push apart the objects and prevent overlap occurring beyond the blur region. The advantage of formulating the collision interaction as a stress instead of a force-pair is that it immediately ensures the conservation of momentum numerically. In particular, for a pair of solids $(i)$ and $(j)$, the collision stress is defined as
\bq
\vsigma_\text{col}=-\eta\min\{f(\phi^{(i)}),f(\phi^{(j)})\}(G^{(i)}+G^{(j)})(\vn\otimes\vn-\tfrac12{\mathbf{1}})\label{eq:coll}
\eq
where $\vn$ is a unit vector field defined by
\bq
\vn=\displaystyle\frac{\nabla(\phi^{(i)}-\phi^{(j)})}{\|\nabla(\phi^{(i)}-\phi^{(j)})\|_2}.
\eq
Note this vector $\vn$ is perpendicular to the midsurface that is defined by $\phi^{(i)}=\phi^{(j)}$, since a point on that surface is equidistant between the two solid boundaries~\cite{valkov2015eulerian}. $\eta$ is a dimensionless constant, $G^{(i)}$ is the object-dependent shear modulus of object $(i)$,  and $f(\phi)$ is a function that is zero when $\phi\geq\epsilon$ and grows to 1 as $\phi$ approaches $-\epsilon$. In the rare case where the edge is within three or more solid blur zones, the calculation is repeated to find a $\vsigma_{\text{col}}$ for each pair. We also note that the shear modulus $G^{(i)}$ can be replaced with other material quantities, with the property that $\vsigma_{\text{col}}$ becomes larger as the structure is stiffer. The method is not sensitive to the exact functional form of $f$, but here we use
\bq
f(x)=\left\{\begin{array}{lr}1&\qquad \text{for $x\leq -\epsilon$,}\\ \frac{1}{2}(1-\frac{x}{\epsilon})& \qquad \text{for $x<\epsilon$,}\\
0&\qquad \text{for $x\geq\epsilon$.}\end{array}\right.\label{eq:fscale}
\eq
which is the same as in previous work. Since the collision stress is incorporated into the system before the projection step, it does not affect the pressure and rigid stress correction after the projection. Therefore, it is natural to adopt the same multi-body contact model for soft bodies, rigid--rigid interactions, and soft--rigid interactions but with a few modifications. First, for rigid bodies, the shear modulus $G$ is infinity in theory, which is infeasible to use in the computation. Instead, we derive a collision shear modulus for rigid structures to be used in Eq.~\eqref{eq:coll} based on dimensional analysis~\cite{barenblatt_scaling} so the magnitude of the collision stress is large enough to push apart objects as described next.

We consider a rigid body of density $\rho$ with characteristic length $\mathcal{L}$ moving in a fluid domain with characteristic length $\mathcal{H}$. For example, in a fluid box, $\mathcal{H}$ could be the length of the maximum edge. The object initially has velocity $\vu_0$, and is either driven by a body force with acceleration $\va$, or driven by the fluid field with characteristic velocity $U$. When two objects collide, the impact energy per length in the third dimension scales as
\bq
E_{\text{col}}\sim \frac{1}{2}\rho {\mathcal{L}}^2 |\vu_0|^2+\rho {\mathcal{L}}^2\|\va\|{\mathcal{H}}\quad \text{or}\quad
E_{\text{col}}\sim \frac{1}{2}\rho {\mathcal{L}}^2 (|\vu_0|^2+U^2).
\eq
We then define the collision shear modulus for a rigid structure as
\bq
G_{\text{col}}=\zeta\frac{E_{\text{col}}}{\epsilon^2},
\eq
where $\zeta$ is a dimensionless constant.  $G_{\text{col}}$ is used as a rigid body's shear modulus in Eq.~\eqref{eq:coll} for the purposes of constructing the repulsion stress. This functional form is chosen so that the collision energy $E_{\text{col}}$ is approximately balanced by the work done by the repulsive stress field as the objects approach each other in the region where the collision stress is activated. 

According to Eq.~\eqref{eq:coll}, a collision stress is exerted in the domain $-\epsilon<\phi<\epsilon$. For soft objects, this is exactly the same domain as the blurred zone where the deformation stress of the structure exists. The extra stress incites both fluid and structural response. For rigid objects, the rigid stress is only computed in the domain $-\epsilon<\phi<0$. If we apply the collision stress in $0<\phi<\epsilon$, the fluid domain will be affected by the extra stress, which leads to nonphysical fluid squeezing motion in the contact zone. To avoid this phenomenon, when both approaching objects are rigid, we only apply the divergence of contact stress $\nabla\cdot\vsigma_{\text{col}}$, {\it{i.e.}}, the repulsive force field, in the rigid domain $\phi<0$. Any momentum imbalance from this change is small and tends to zero as $\epsilon \to 0$.\footnote{This issue could be eliminated entirely by computing the total repulsive force on each rigid object and applying a global rescaling to ensure equal and opposite forces, but we do not consider this here.}
\subsection{Extra viscosity and stability analysis}
After the projection step, the velocity field in the rigid domain is shear-free, while the velocity directly adjacent to the object in the fluid domain only satisfies the incompressibility condition. Therefore, there is a velocity gradient jump in grid points adjacent to the fluid--structure interface, which could lead to a kink of the velocity field near the interface. To rectify this, we incorporate an extra artificial viscous stress inside the blur zone, which is added to the stress of the solid. Based on dimensional considerations, the artificial viscosity should satisfy
\bq
\mu_e = \kappa_e\sqrt{G\rho}\max\{\Delta x,\Delta y\}
\eq
where $\kappa_e$ is a dimensionless constant. This is the same formula we used in previous fluid--soft interactions~\cite{rycroft2018reference}, except that for rigid objects, the shear modulus $G$ is based on the collision strategy instead of the true solid shear modulus. The artificial viscosity only exists in the blur zone and within the structure, and takes a different value for each object. To help stabilize the numerical system, we multiply the stress with a smoothed delta function (given by the derivative of the smoothed Heaviside function) that amplifies the effect of viscous damping near the fluid--rigid interface. We use
\bq
\vsigma_e = \mu_e (1-H(\phi(\vx)))( 1 + q \epsilon H'_\epsilon(\phi)) \nabla \vu \label{eq:visdamping}
\eq
where $q$ is a dimensionless constant. Based on a variety of tests in this work and in previous work~\cite{rycroft2018reference}, we use $q=1$ and $\kappa_e=0.4$ throughout the results presented in this paper, which is the same artificial viscosity used in fluid--soft simulation to be consistent. When computing $\vu^*$, the extra viscous stress $\vsigma_e$ will be included in $\vsigma_d^n$ and $\vsigma_r^n$, and thus does not affect the linear matrix in the projection step. We note that the extra viscosity remains constant as grid size shrinks in the fluid--rigid interface. However, the overall region where the extra viscous stress is nonzero shrinks with the grid size; any errors that are introduced by the extra viscous stress vanish as the grid is refined.

Therefore, in a simulation with fluid density $\rho_f$ and viscosity $\mu_f$, soft solid density $\rho_s$, soft object shear modulus $G_s$, rigid solid density $\rho_r$, and collision shear modulus $G_\text{col}$, the time step needs to satisfy the following stability restrictions:
\begin{enumerate}
    \item Fluid viscous stress constraint: $\Delta t_1 = \displaystyle\frac{\rho_f}{2\mu_f(\Delta x^{-2}+\Delta y^{-2})}$;
    \item Solid wave speed constraint: $\Delta t_2=\min\left\{\displaystyle\sqrt{\frac{\rho_s}{G_s}},\sqrt{\frac{\rho_r}{G_\text{col}}}\right\}\min\{\Delta x,\Delta y\}$;
    \item Extra viscous stress constraint: $\Delta t_3=\displaystyle\frac{\min\{\rho_r,\rho_s\}}{2\mu_e(\Delta x^{-2}+\Delta y^{-2})}$.
\end{enumerate}
The second time constraint comes from the CFL condition for solids, since the shear wave speed in the solids is $c=\sqrt{G/\rho}$. The time step $\Delta t$ is chosen to be smaller than the minimum of the above three conditions with an extra padding factor, so that
\bq
\Delta t=\min\{\alpha_{\text{pad}}\Delta t_1,\beta_{\text{pad}}\Delta t_2, \gamma_{\text{pad}}\Delta t_3\}.
\eq
As commented by previous work~\cite{rycroft2018reference}, the first two restrictions arise from the physical stresses and the last one is for the artificial stress. We use $\alpha_{\text{pad}}=\beta_{\text{pad}}=0.4$, and $\gamma_{\text{pad}}=0.8$ so that the timesteps arising from the physical terms are applied more stringently than the timestep from the artificial stress.

In Appendix \ref{sec:conv-test}, we present a detailed convergence study and show that our method can achieve a second-order convergence rate under certain assumptions. Since we use a sharp interface for stress terms, it is not surprising that the largest error happens near the fluid-solid interface.
\section{Results and Discussion}
\label{sec:results}
For simplicity, we only consider the case of uniform structure density $\rho_s$ or $\rho_r$ in this work. However, we note that the model can be easily generalized to density-varying solids. Our results also focus on the
case of equal grid spacing, $\Delta x = \Delta y=h$.
\subsection{Falling rigid object}
We first consider the classical problem of a rigid object falling in a fluid channel. In the case where the object has a large aspect ratio that is perpendicular to the direction of motion, this problem can be treated as two-dimensional. In the low Reynolds number regime, the terminal velocity of a falling cylinder with radius $r$ in an infinitely long channel with width $2L$ has been found as~\cite{wang2004extended,wang2008immersed,gibou2012efficient,happel2012low}
\bq
V=\displaystyle\frac{(\rho_r-\rho_f)gr^2}{4\mu_f}\left(-\ln\left(\frac{r}{L}\right)-0.9157+ 1.7244\left(\frac{r}{L}\right)^2-1.7302\left(\frac{r}{L}\right)^4\right). \label{eq:analfall}
\eq
We simulate a fluid domain of $2L\times 8L$, which matches the size of channel used by Gibou and Min~\cite{gibou2012efficient}, and we take $L=\SI{1}{cm}$ and $r=\SI{0.3}{cm}$. A gravitational acceleration of $g=\SI{500}{cm/s^2}$ is applied in the negative $y$ direction. Here, we take the Reynolds number as $\Rey=\rho_fVd/\mu_f$ with the analytical terminal velocity $V$ to be the characteristic velocity and $d=2r$ to be the characteristic length. No-slip and no-penetration boundary conditions are applied at the four walls. The effect of gravity is implemented as an external body force that is only applied to the rigid structure not the fluid, and the buoyancy is captured by modifying the body force to be $(\rho_r-\rho_f)g$.

In Fig.~\ref{fig:dropanal}, we consider $\rho_f=\SI{1.0}{\gram/\cm^3}$ and $\rho_r=\SI{2.0}{\gram/\cm^3}$, with $\mu_f=\SI{0.1}{\pascal\second}$ and $\mu_f=\SI{0.2}{\pascal\second}$, respectively. We compare the numerical terminal velocity we obtain with different grid resolutions with the analytical solution and show the dimensionless results. The velocity is scaled by $\sqrt{dg}$ and the time is scaled by $\sqrt{d/g}$ where $d=\SI{0.6}{cm}$ is the diameter of the circle. The Reynolds number is $\Rey\approx 2.9$ and $\Rey\approx 0.73$ for the two viscosity values, respectively. The analytical formula in Eq.~\eqref{eq:analfall} is derived under the Stokes assumption for an infinitely long channel. Therefore, it is not surprising to see the deviation of our results with the analytical formula for larger $\Rey$ due to increasing inertia effects, which is consistent with the observation in others' work~\cite{gibou2012efficient,robinson2011symmetric}. We note that as $\Rey$ becomes smaller, the numerical solution gets closer to the analytical formula. We also note that for coarser resolutions $50\times 200$ and $100\times 400$, there are some numerical oscillations obtained in the solution due to the fluid-rigid sharp interface. This issue is resolved when the resolution is fine enough.

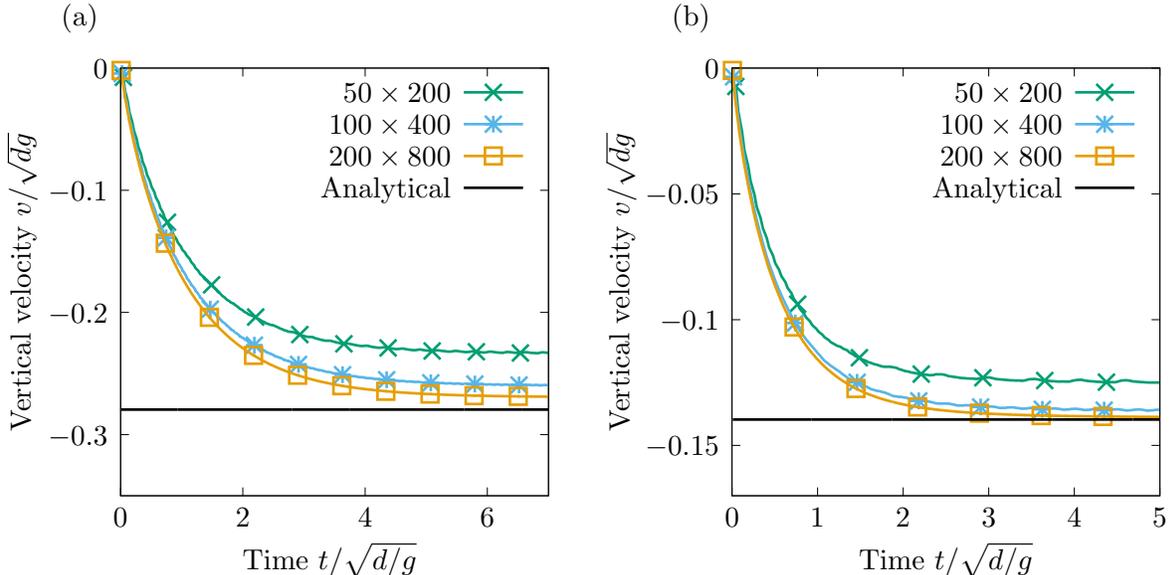
\Eoneplots{visc}{\label{fig:dropanal} The velocity history of a falling rigid object using different simulation resolutions, for a fluid viscosity of (a) $\mu_f=\SI{0.1}{\pascal\second}$ and (b) $\mu_f=\SI{0.2}{\pascal\second}$. Other physical parameters used here are $\rho_f=\SI{1.0}{\gram/\cm^3}$ and $\rho_r=\SI{2.0}{\gram/\cm^3}$.}

Since the fluid viscous term $\vsigma_f$ is treated explicitly in our method and the time step $\Delta t$ scales as $1/\mu_f$, our method is more suitable for problems with moderate Reynolds number instead of near-Stokes flow. Next, we show examples in the moderate $\Rey$ regime. For comparison purposes, we compare the results obtained by the fluid--rigid solver with the results obtained by treating the solid as hyperelastic with a gradually increasing shear modulus $G$. We note here that in our previous work \cite{rycroft2018reference}, we have performed a thorough investigation on the validation of our fluid-soft interaction solver by comparing with other numerical and experimental results. Its accuracy was validated spanning over a range of density ratios, shear modulus values, and different geometries. 

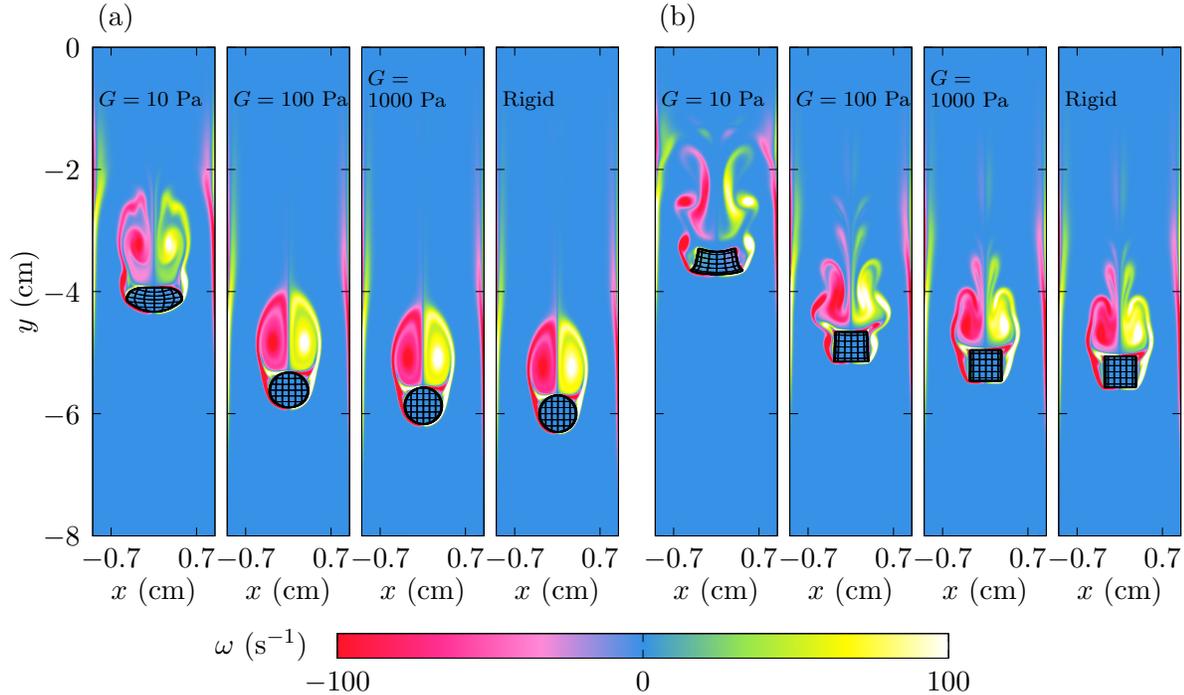
\Eoneplots{single-drop}{\label{fig:circledrop1} Instantaneous vorticity $\omega$ of a free-falling object in a channel obtained at $t=\SI{0.35}{s}$. Common simulation parameters are $(\rho_f,\rho_s,\rho_r)=(1.0,2.0,2.0)$\,\si{\gram/\cm^3}, $\mu_f=\SI{1e-3}{\pascal\second}$ and $g=\SI{500}{\cm/s^2}$. Rigid solutions are compared with soft solutions obtained with $G=\SI{10}{\pascal},\SI{100}{\pascal},\SI{1000}{\pascal}$. The thick black line marks the fluid--structure interface. The thin lines inside the structure are contours of the components of the reference map and indicate the level of deformation. (a) Falling circle comparison. (b) Falling square comparison.}

In Fig.~\ref{fig:circledrop1}, we show snapshots of vorticity $\omega$ of free-falling objects in a channel. Panel (a) shows the vorticity of a falling circle and panel (b) shows the results of a falling square. All simulation results are obtained at $t=\SI{0.35}{s}$ with resolution $200\times800$. The vorticity $\omega=\partial_x v-\partial_y u$ is computed on each grid cell corner, using central finite differences of the velocities in the four adjoining grid cell centers. The rigid results are compared with the behavior of increasingly stiff soft simulations with $G$ increasing from $\SI{10}{\pascal}$ to $\SI{1000}{\pascal}$. All simulations use $(\rho_f,\rho_s,\rho_r)=(1,0,2.0,2.0)$\,\si{\gram/\cm^3}, $g=\SI{500}{\cm/s^2}$, and $\mu_f = \SI{1e-3}{\pascal\second}$. The fluid--structure interfaces are indicated with thick lines, while the thin lines inside the structures are contours of the components of the reference map. They indicate the level of deformation for a structure and remain a square grid when the object is rigid.

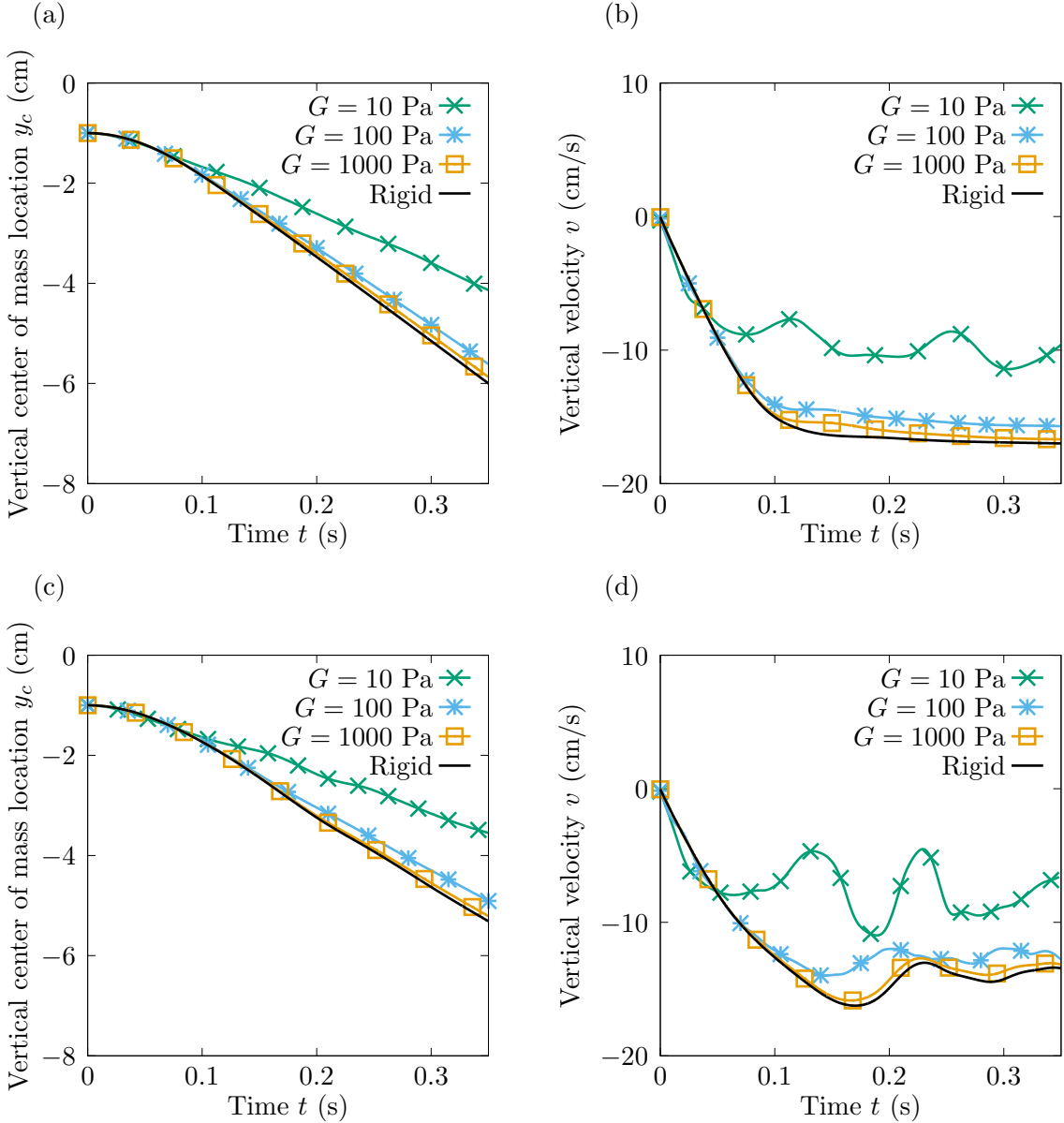
\Eoneplots{singleobj}{\label{fig:droptraj}Trajectories of the vertical location and velocity for the center of mass of the objects corresponding to the simulation in figure \ref{fig:circledrop1}. Panels (a) and (b) show the results for falling circles and panels (c) and (d) show the results for falling squares. }

In Fig.~\ref{fig:droptraj}, we show the corresponding vertical velocity $v$ and vertical center of mass location $y_c$ of the above simulations. Panels (a) and (b) show the results for circles and panels (c) and (d) show the results for squares. For both cases, significant solid deformation was observed in objects with smaller shear modulus ($G=10\si{\pascal}$) due to the stress on the solid boundary from the fluid interaction. As the objects become stiffer, their internal deformation decreases. The rigid solution matches well with the trend of the soft solutions as $G$ increases, with the rigid solutions falling a bit faster than the soft solutions for the highest value of $G=\SI{1000}{\pascal}$. We emphasize here that one advantage of our direct fluid-rigid solver, compared to approximating rigid objects as very stiff ones, is that it allows a much larger time step. For this example, when $G=1000\si{\pascal}$, the main time constraint for the fluid-soft solver comes from the CFL constraint for solid shear waves, giving a time step of $\Delta t=9.5\times 10^{-5}\si{\second}$. For the rigid solver, the main time constraint comes from the fluid viscosity and extra viscosity, giving a time step of $\Delta t= 1.1\times 10^{-3}\si{\second}$. 

A benefit of our current model is that it can handle solids within a large density range without any modifications. To demonstrate this, we present a falling-circle example with a density ratio $\rho_r/\rho_f$ ranging from 0.1 to 10. A circle with radius $r=0.3\si{\cm}$ was placed in the middle of a channel of $[-1, 1]\times [-6, 0]\,\si{\cm^2}$. The circle was initially placed at $y=-3$ with an acceleration of $g=\SI{500}{\cm/s^2}$ in the negative direction. Common simulation parameters are $\rho_f=1.0$ \si{\gram/\cm^3} and $\mu_f=\SI{1e-3}{\pascal\second}$. In figure \ref{fig:densitydiff}, snapshots of vorticity are shown at $t=0.1\si{\s}$. As we expect, the objects sink faster when denser, and float upwards when the density of the solid is less than the fluid. 
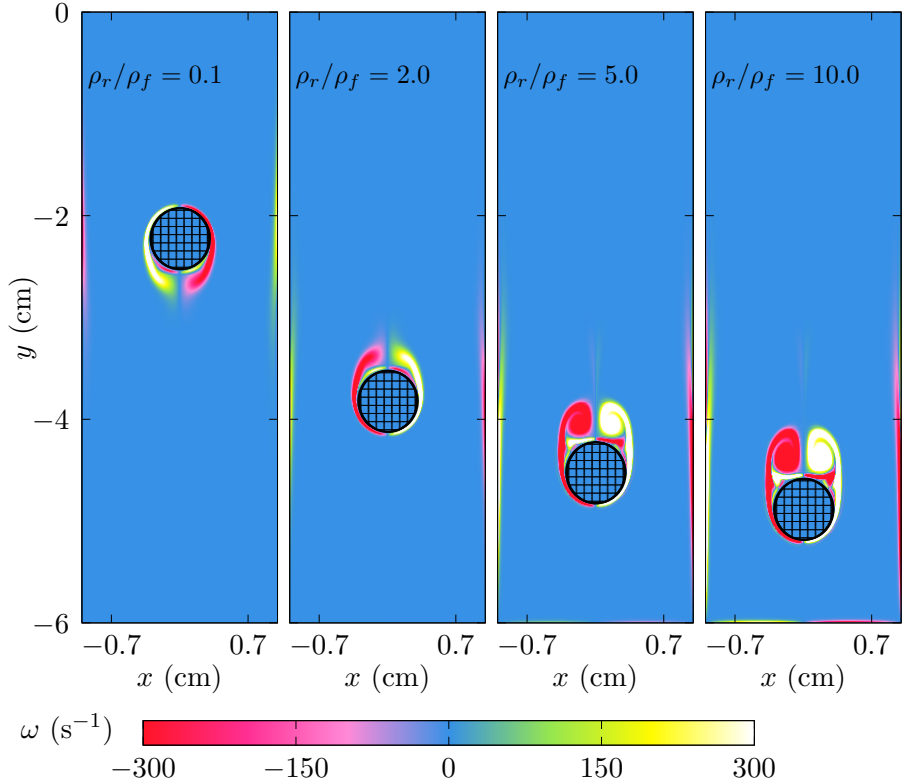
\Eoneplots{densitydiff}{\label{fig:densitydiff}Instantaneous vorticity $\omega$ of a free-falling circle with different density ratio in a channel obtained at $t=\SI{0.1}{s}$. Common simulation parameters are $\rho_f=1.0,2.0,2.0$\,\si{\gram/\cm^3}, $\mu_f=\SI{1e-3}{\pascal\second}$ and $g=\SI{500}{\cm/s^2}$. The density ratio of solid and fluid varies from 0.1 to 1, with an acceleration in the negative $y$-direction. The thick black line marks the fluid--structure interface. The thin lines inside the structure are contours of the components of the reference map and indicate the level of deformation. }
\subsection{Falling soft object with rigid components}
With a simple modification, the method also admits a straightforward approach for simulating soft structures with rigid components inside. To do so, we need to construct another function that defines the soft--rigid interface. This can be done with another level-set function that defines the shape of the rigid domain, which is evaluated through the reference map values of the rigid domain. The reference map values in the rigid domain are updated with Eq.~\eqref{eq:updateref} through the rigid kernel center of mass motion. This is useful particularly when the rigid components have complicated geometries, and we can take advantage of the level-set function to ensure an accurate description of the soft--rigid interface. However, in the case where the rigid component has a simpler geometry, we can use a simple calculation to determine the rigid points.

In the following example, we consider a case with inner circular rigid components with radius $R$ and center $(x_\text{rc}(t),y_\text{rc}(t))$. Any grid point $(x,y)$ is assigned rigid when its distance to the center $r=\sqrt{((x-x_\text{rc})^2+(y-y_\text{rc})^2}$ is $\leq R$. Similar to the fluid--rigid interaction, there is a velocity gradient jump at the soft--rigid interface since the rigid stress correction is only updated in the pure rigid domain. To stabilize the system, we again add an artificial viscous damping using the same formula as Eq.~\eqref{eq:visdamping} in the domain $r<R+\epsilon$.

In Fig.~\ref{fig:rod-drop}, we show the example of an elastic rod with rounded ends sedimenting in a fluid box of $[-1,1]\times[-3,0]$\,\si{\cm^2} with a negative $y$-direction acceleration of $g=\SI{500}{\cm/s^2}$. The rod has a length of $\SI{1.25}{\cm}$ and a width of $\SI{0.5}{\cm}$ with shear modulus $G=\SI{10}{\pascal}$. Two circular rigid components within the rod are defined initially at $(-0.4,-0.5)$\,\si{\cm} and $(0.4,-0.5)$\,\si{\cm} with radius $\SI{0.1}{\cm}$.

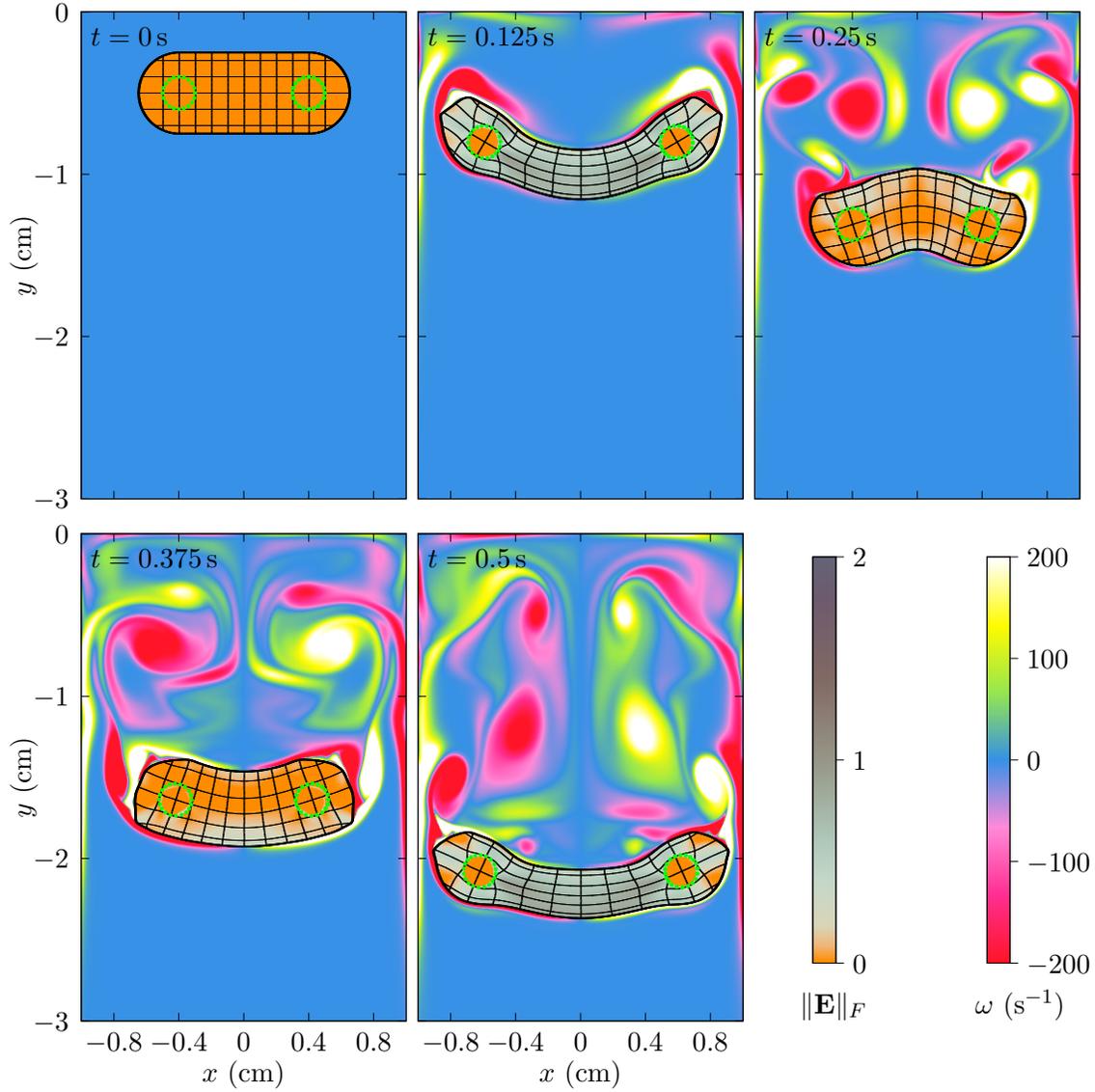
\Eoneplots{rod-drop}{\label{fig:rod-drop}Snapshots of a simulation where mixed soft--rigid rod falls in a fluid-filled box, showing vorticity $\omega$ in the fluid and the Frobenius norm of Hencky strain $\|\mathbf{E}\|_F$ in the rod. The rod has two inner rigid circular areas at $(\pm0.4,-0.5)\,\si{\cm}$ with radius $\SI{0.1}{\cm}$. The rest of the rod is made of soft material with shear modulus $G=\SI{10}{\pascal}$. The thick lines mark the fluid--structure interfaces. The thin lines inside the structure are contours of the components of the reference map and indicate how the structure is deformed. The soft--rigid interface is shown with green dots. The other simulation parameters are $(\rho_f,\rho_s,\rho_r)=(1.0,2.0,2.0)\,\si{\gram/\cm^3}$, and $\mu_f=\SI{1e-3}{\pascal\second}$.}

In Fig.~\ref{fig:rod-drop}, we plot the vorticity field in the fluid domain, and the Frobenius norm of the Hencky strain $\|\mathbf{E}\|_F$ inside the object where $\mathbf{E} = \log(\sqrt{\mathbf{F}\mathbf{F}^\trans})$ with $\mathbf{F}$ as the deformation gradient tensor~\cite{hencky1928uber}. The strain variable is computed at each grid cell corner using finite center difference of the reference map values $\vxi$ from the adjoining four grid cell centers. For rigid structures, this value remains zero during the simulation, and the soft--rigid interface is shown with green dots. The thin dashed lines inside the object are the contours of the components of the reference map, which also indicates how the structure is deformed. In the rigid part, the two families of contours indeed move rigidly, while the rest of the body has finite deformations during the falling.

\subsection{Multi-body contact}
\subsubsection{Drafting-kissing-tumbling case}
We consider two rigid circles with identical density $\rho_r$ and radius $r$ accelerating from rest due to the action of gravity in the negative $y$-direction with acceleration $g=\SI{981}{\cm/s^2}$. Initially, they have the same horizontal position. The upper circle gradually approaches the lower one due to the reduced drag formed by the vortex wake of the lower circle. After the first collision (``kissing"), a ``tumbling" stage emerges with instabilities, followed by the ``drafting" stage where horizontal movement is observed. This drafting-kissing-tumbling (DKT) case has been discussed frequently in the fluid--rigid interaction literature~\cite{uhlmann2005immersed,hu2001direct,glowinski2001fictitious,feng2004immersed,wang2008immersed} where different collision models were implemented. In particular, we compare our result quantitatively with the results obtained by Uhlmann~\cite{uhlmann2005immersed} and Glowinski \etal \cite{glowinski2001fictitious}. The former work is based on immersed boundary method with direct forcing, while the latter one is based on a Lagrange-multiplier-based fictitious domain methods. We choose boundary conditions and resolutions consistent with these studies. The simulation parameters used in this work are summarized in Table \ref{tab:dkt}.

\begin{table}[ht]
  \caption{\label{tab:dkt}Dimensional parameters used in the drafting-kissing-tumbling (DKT) simulation}
\begin{tabular}{p{8cm}llp{8cm}}
\hline\hline
Domain: $[-1, 1]\times [-6, 0]\,\si{\cm^2}$&Circle radius: $r=\SI{0.125}{\cm}$\\ \hline
Gravitational acceleration $g=\SI{981}{\cm/s^2}$ &Fluid viscosity: $\mu_f=\SI{0.01}{\pascal\second}$\\ \hline
 \multicolumn{2}{l}{Initial location: $\vx^{(1)}_c(0)=(0,-1)\,\si{\cm}$, $\vx_c^{(2)}(0)=(0,-1.5)\,\si{\cm}$}\\ \hline
 Density:  $(\rho^{(1)}_r,\rho_r^{(2)},\rho_f)=(1.5,1.5,1.0)\,\si{\gram/\cm^3}$ &  $\Delta x=\Delta y=1/\SI{256}{cm}$\\ \hline\hline
\end{tabular}
\end{table}

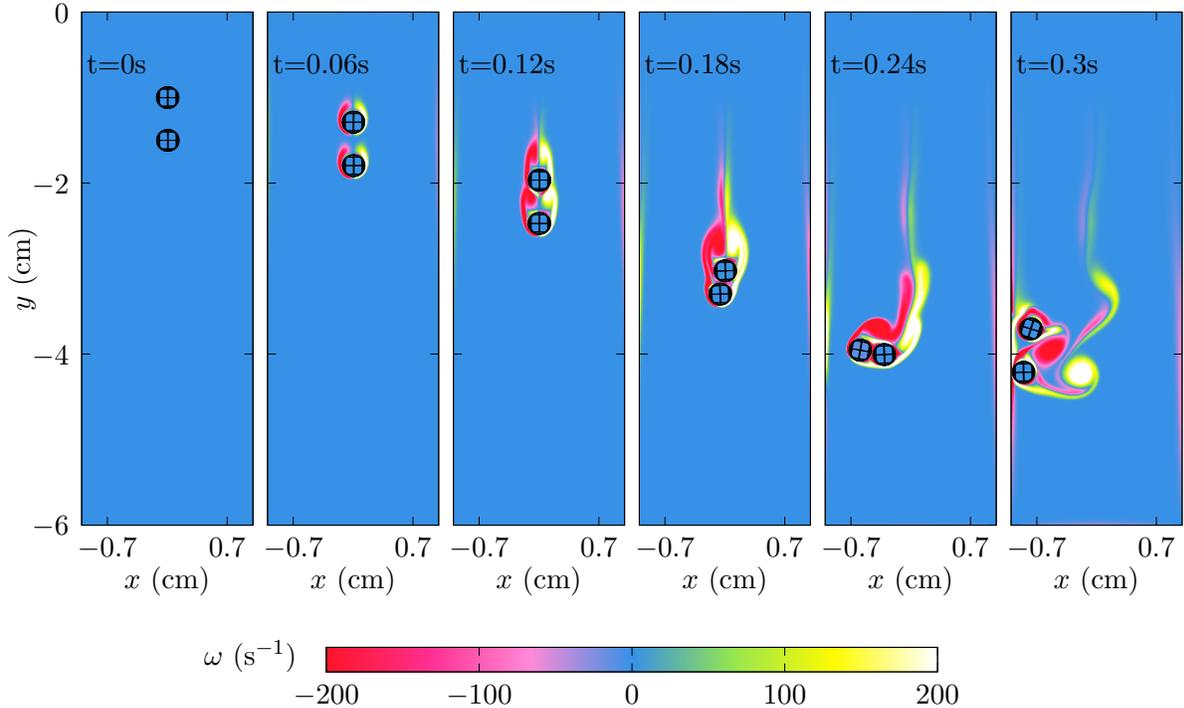
\Eoneplots{dkt}{\label{fig:dktvor}Snapshots of vorticity $\omega$ in a simulation of drafting-kissing-tumbling of two rigid circles sedimenting in a fluid box of $[-1,1]\times[-6,0]\,\si{\cm^2}$ at $t=\SI{0}{s},\SI{0.06}{s},\SI{0.12}{s},\SI{0.18}{s},\SI{0.24}{s},\SI{0.3}{s}$. The collision shear modulus used is $G_\text{col}=\SI{40000}{\pascal}$.}
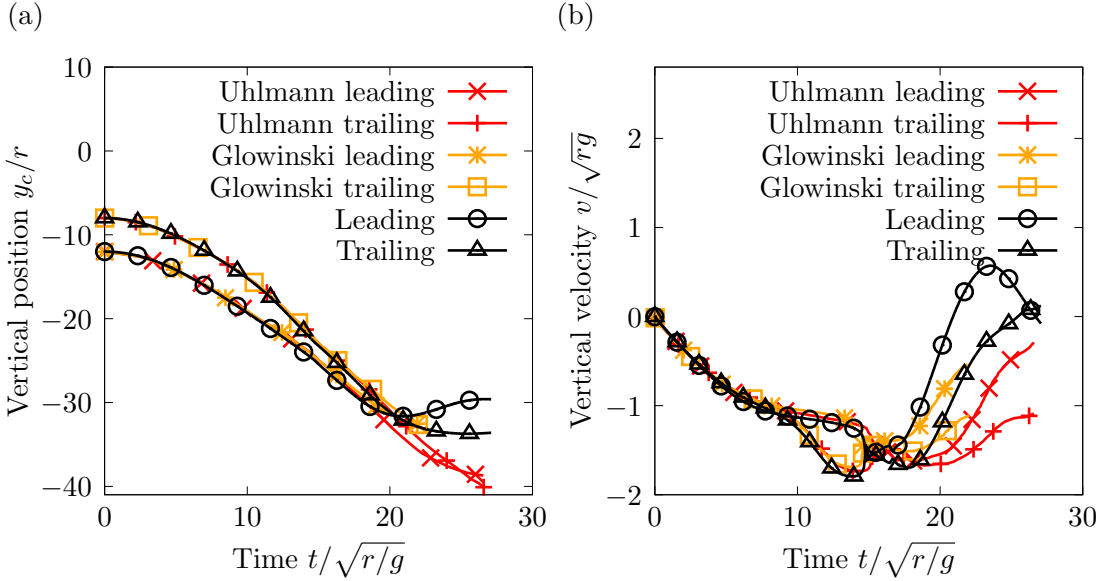
\Eoneplotsw{dkt-traj}{\label{fig:dkt-traj}Time trajectories of dimensionless vertical location $y_c/r$ (a) and velocity $v/\sqrt{rg}$ (b) for the center of mass of the objects in the drafting-kissing-tumbling simulation. The leading object is initially located at $y=-\SI{1.5}{\cm}$, and the trailing object is initially located at $y=-\SI{1}{\cm}$. }

In Fig.~\ref{fig:dktvor}, we show snapshots of vorticity throughout the simulation. In Fig.~\ref{fig:dkt-traj}, we show a comparison of the center of mass vertical location $y_c$ and the vertical velocity $v$ obtained with our model, compared with the data shown by Uhlmann~\cite{uhlmann2005immersed} and Glowinski \etal \cite{glowinski2001fictitious}. We nondimensionalize different simulation results by circle radius $r$ and gravitational acceleration $g$ for comparison. Uhlmann~\cite{uhlmann2005immersed} added the acceleration in the positive $x$-direction and we made a corresponding modification to this data to match our simulation settings. Since we used a blurring technique on the density over the fluid-rigid interface, we also adjusted the acceleration so that the total body force on the object is the same with the other simulations. We observed an agreement with both sets of data before the first collision: on panel (b) this time is indicated by the sudden drop of the leading velocity caused by the collision at around $t^*=t/\sqrt{r/g}=15$. After the collision, we found that in the tumbling stage, two objects eventually switch their vertical positions and the trailing object becomes the leading one, as shown in Fig.~\ref{fig:dkt-traj}(a) where the leading and trailing curve come cross each other. This is consistent with the other simulation results.

However, there is not a qualitative comparison among simulations with different numerical methods in the drafting and tumbling stage. As discussed by Fortes \etal\cite{fortes1987nonlinear}, the DKT phenomena is essentially a breakup of the particle
positions in an unstable configuration, thus an exact agreement after kissing may not be expected. 
In Fig.~\ref{fig:dkt-trajx}, we plot the magnitude of the horizontal displacement and velocity for the center of mass. The horizontal drafting process, which is also observed in other simulations, is quite clear as both objects move horizontally until they hit the boundary.
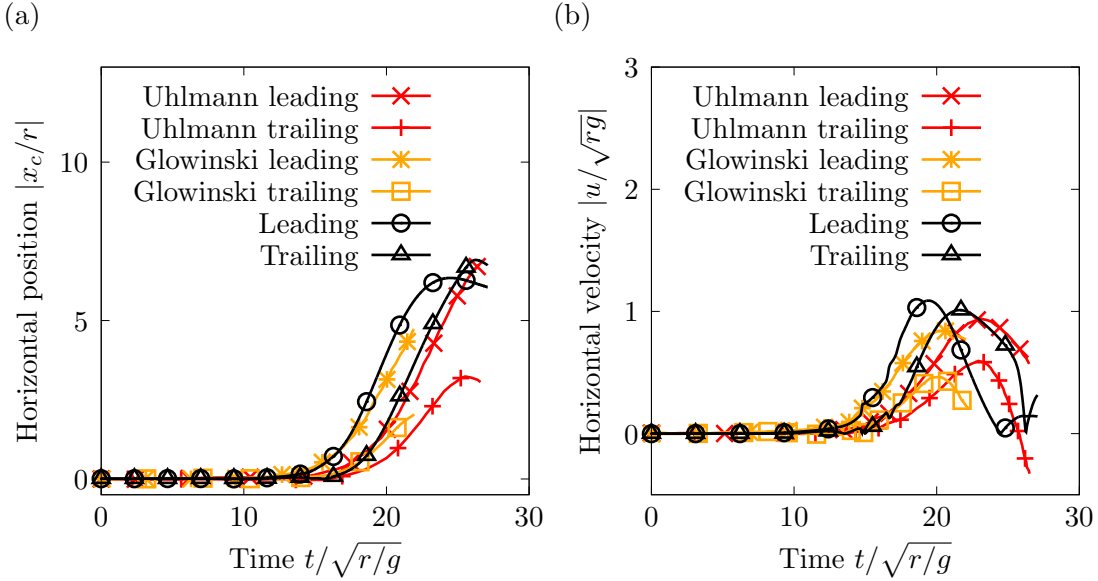
\Eoneplotsw{dkt-trajx}{\label{fig:dkt-trajx}Trajectories of magnitude of horizontal location $x_c$ (a) and velocity $u$ (b) for the center of mass of the objects in the drafting-kissing-tumbling simulation. The leading object is initially located at $y=-\SI{1.5}{\cm}$, and the trailing object is initially located at $y=-\SI{1}{\cm}$.} 

This example also demonstrates our wall-contacting strategy when the object is close to a wall boundary. When the object is within $w=3\epsilon$ of the wall-boundary, a direct repulsive force $\mathbf{F}_w$ will be added when computing the intermediate step solution $\vu^*$ to repel the object away. It is formulated as: 
\bq
\mathbf{F}_w=\frac{k_\text{rep}}{w}\delta \left(\displaystyle\frac{\|\vx-\vx_w\|}{w}\right)f(\phi(\vx))
\eq
where $k_\text{rep}=20w$ is a repulsive coefficient that determines the strength of the repulsion, $\delta(x)=\displaystyle\frac{1}{2\epsilon}\left(1+\cos\frac{\pi x}{\epsilon}\right)$ is a smoothed delta function, and $f(x)$ is the scaling function that goes from 0 to 1 when $x$ goes from $\epsilon$ to $-\epsilon$, as defined in Eq.~\eqref{eq:fscale}.

\subsubsection{Mixed soft and rigid interactions}\label{section:multibody}
Since the current projection method for implementing fluid--rigid interactions uses the same basic framework  as that for incompressible fluid--soft interactions~\cite{rycroft2018reference}, it is straightforward to combine both features into one simulation. We use the collision strategy discussed in Subsection \ref{sub:multi_c} . 

Here we present an example to illustrate the capacity of our method in dealing with mixed soft and rigid interactions. In a non-periodic box $\Omega=[-1,1]\times[-1,1]\,\si{\cm^2}$ with fluid density $\rho_f=\SI{1.0}{\gram/\cm^3}$ and dynamic viscosity $\mu_f=\SI{0.001}{\pascal\second}$, sixteen squares with density $\rho_s=\rho_r=\SI{2.0}{\gram/\cm^3}$ are inserted at random positions in the box, with side lengths chosen uniformly over the range $[0.1,0.4]\,\si{\cm}$, as shown in Fig.~\ref{fig:multisquare}. Among the sixteen squares, four of them are rigid and the rest are soft with shear modulus $G=\SI{50}{\pascal}$. Initially, all the soft squares are set with an initial angular velocity chosen uniformly from the range $\SI{-50}{rad/s}$ to $\SI{50}{rad/s}$, and all the rigid squares are stationary. A gravitational acceleration $g$ of $\SI{500}{\cm/s^2}$ in the negative $y$-direction is applied, so that the squares sediment at the bottom of the box. The resolution for this simulation is $500\times500$.

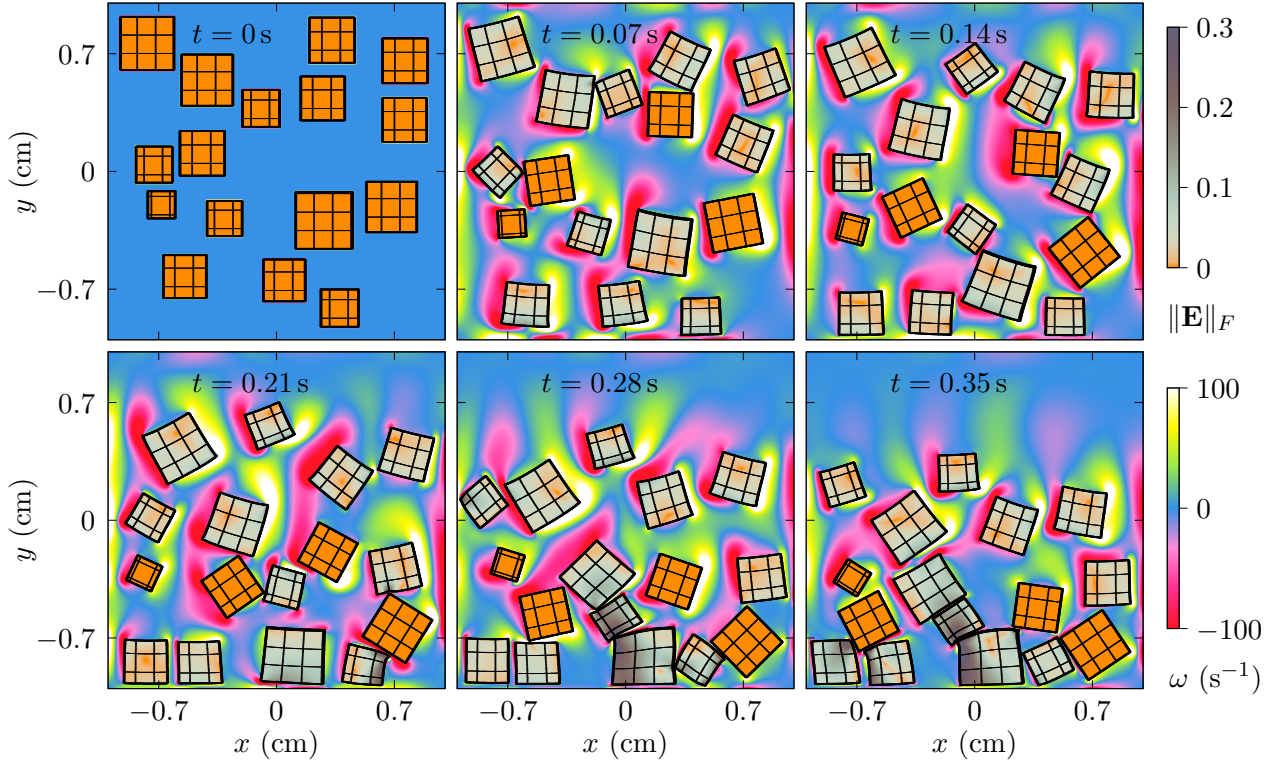
\Eoneplots{multi-drop}{\label{fig:multisquare}Snapshots of vorticity $\omega$ and the Frobenius norm of the Hencky strain $\|\mathbf{E}\|_F$ in a simulation of a mixed soft and rigid squares sedimenting in a fluid-filled box. Sixteen squares are inserted at random positions in the box, four of which are rigid and the rest are elastic structures with shear modulus $G=\SI{50}{\pascal}$. The thick lines mark the fluid--structure interfaces. The thin lines inside the structure are contours of the components of the reference map and indicate how the structure is deformed. The other simulation parameters are $(\rho_f,\rho_s,\rho_r)=(1.0,2.0,2.0)\,\si{\gram/\cm^3}$, $\mu_f=\SI{0.01}{\pascal\second}$ and $g=\SI{500}{\cm/s^2}$. }

In Fig.~\ref{fig:multisquare}, we show the vorticity field $\omega$ in the fluid domain and plot the Frobenius norm of the Hencky strain, $\|\mathbf{E}\|_F$, in the solid domain. For rigid structures, this value remains zero during the simulation as expected. During the simulation, soft--soft interaction, soft--rigid interaction, and fluid--structure interaction are all well-captured by our model.

\subsection{Kinematic boundary conditions}
We now consider extending our method to apply kinematic control to the structure, which is a typical requirement in experimental and modeling scenarios. In fluid--rigid interactions, we implement the rigidity by formulating extra constraints that the velocity field needs to satisfy at the projection step. Similarly, we can compute extra body forces to enforce kinematic boundary conditions simultaneously with incompressibility and rigidity constraints in a single projection step. To illustrate our approach, we first consider fully prescribing the motion of a rigid object. This can be defined
by a translation velocity $\vu_T=(u_T,v_T)^\trans$ and an angular velocity $\boldsymbol\lambda = (0,0,\lambda)^\trans$. The velocity field is then given by
\bq
\vu(\vx,t) =\vu_T+\vlambda\times(\vx-\vx_c(t))
\eq
where $\vx_c(t)$ is the center of mass. Define $\vecr= \vx-\vx_c(t) = (r^{(x)},r^{(y)})^\trans$, and then we can define
two extra body forces 
\bq
\mathbf{f}_T = \left(\begin{array}{c}
     f^{(x)} \\
     f^{(y)}
\end{array}\right), \qquad \mathbf{f}_R(x,y)=\boldsymbol\Lambda\times\vecr=\left(\begin{array}{c}
     -\Lambda r^{(y)} \\
     \Lambda r^{(x)}
\end{array}\right)
\eq
to constrain the translational motion and rotational motion, respectively. We note that $f^{(x)},f^{(y)},$ and $\boldsymbol\Lambda=(0,0,\Lambda)$ are uniform across the rigid grid points. Compared to the regular fluid--rigid solver, only three extra unknowns are added into the system. Now, the velocity update formula Eq.~\eqref{eq:updatev} becomes
\bq
\vu^{n+1}=\vu^*+\displaystyle\frac{\Delta t}{\rho(\phi^{n+1/2})}\left(
-\nabla p^{n+1}+\nabla\cdot\vsigma_r^{n+1}+\mathbf{f}_T+\mathbf{f}_R,\right).
\eq
In the projection step, the velocity field satisfies the incompressibility condition over the whole computational domain, the rigidity condition in the rigid domain, and has the chosen velocities where they are prescribed. Equations \eqref{eq:incomp}, \eqref{eq:dev1} and \eqref{eq:dev2} become
\begin{align}
\nabla\cdot\left(\frac{\Delta t}{\rho(\phi^{n+\frac{1}{2}})}\left(-\nabla p+\left[\begin{array}{c}\theta_x\\-\theta_y\end{array}\right]+\left[\begin{array}{c}\tau_y\\ \tau_x\end{array}\right]+\left[\begin{array}{c}
     f^{(x)} \\
     f^{(y)}
\end{array}\right]+
\Lambda\times\vecr\right)^{n+1}\right) &= -\nabla\cdot\vu^*, \label{eq:incompb}\\
\nabla\cdot\left( \frac{\Delta t}{\rho(\phi^{n+\frac{1}{2}})}\left(\left[\begin{array}{c}p_x\\-p_y\end{array}\right]-\nabla\theta+\left[\begin{array}{c}
     -f^{(x)} \\
     f^{(y)}
\end{array}\right]+\left[\begin{array}{c}
     \Lambda r^{(y)} \\
     \Lambda r^{(x)}
\end{array}\right]\right)^{n+1}\right) &= (u^*_x-v^*_y),\label{eq:dev1b}\\
\nabla\cdot\left( \frac{\Delta t}{\rho(\phi^{n+\frac{1}{2}})}\left(\left[\begin{array}{c}p_y\\p_x\end{array}\right]-\nabla\tau+\left[\begin{array}{c}
     -f^{(y)} \\
     -f^{(x)}
\end{array}\right]+\left[\begin{array}{c}
     \Lambda r^{(x)} \\
     -\Lambda r^{(y)}
\end{array}\right]\right)^{n+1}\right) &= (u^*_y+v^*_x),\label{eq:dev2b}
\end{align}
with extra velocity constraints
\begin{align} 
\frac{\Delta t}{\rho(\phi^{n+\frac{1}{2}})}\left(-\nabla p+\nabla\cdot\left[\begin{array}{cc}
     \theta&\tau \\
     \tau&-\theta
\end{array}\right]+\left[\begin{array}{c}
     f^{(x)} \\
     f^{(y)}
\end{array}\right]\right)&=\vu_T-\vu^*, \label{eq:transv}\\
\frac{\Delta t}{\rho(\phi^{n+\frac{1}{2}})}\left(-\nabla p\times\vecr-\nabla\cdot\left[\begin{array}{c}
     \theta r^{(y)} \\
     \theta r^{(x)}
\end{array}\right]-\nabla\cdot\left[\begin{array}{c}
     \tau r^{(x)} \\
     -\tau r^{(y)}
\end{array}\right]+\Lambda|\vecr|^2\right)&=\lambda\|\vecr\|^2+\vu^*\times\vecr.\label{eq:rotv}
\end{align}
We again make use of a finite-element formulation in the projection procedure. The translation force $\mathbf{f}_T$ and rotation force $\mathbf{f}_R$ are comprised of piecewise bilinear elements. We realized that the discretizations for first order derivatives are skew symmetric. Therefore Eqs.~\eqref{eq:incompb} to \eqref{eq:rotv} again form a symmetric system, and three extra degrees of freedom will be added. 

We note that for fully prescribed fluid--rigid interactions, an alternative approach is to use a spatially-varied force field to constrain the velocity field directly. Our approach instead implements this by constraining the velocity field to be a rigid form with rigid stress and then enforces the prescribed velocity profile with three extra constraint forces. This way of treating the kinematic boundary conditions has two advantages. First, it fully uses the existing fluid--rigid interaction solver and can be easily incorporated into the solver with slight modification. More importantly, this gives us great flexibility in dealing with different kinematic controls. Any combination of horizontal translation velocity, vertical translation velocity, and angular velocity constraints can be implemented through this method. For example, if an object is pinned at a point and allowed to rotate freely in a fluid, then the translation velocity is set to be $\vu_T=(0,0)$ and the body force $\mathbf{f}_T$ will be computed at each step. Only two degrees of freedom will be constrained in this case and $\mathbf{f}_R$ is not computed, which also indicates that Eq.~\eqref{eq:rotv} and all the components involving $\Lambda$ in Eqs.~\eqref{eq:incompb} to \eqref{eq:dev2b} are not included in the linear system. If the object is dragged horizontally in the fluid while being allowed to rotate freely, then only $u_T$ will be constrained and the $x-$component of Eq.~\eqref{eq:transv} will be added into the system. The linear system remains symmetric and the previous  MINRES-QLP method can still be applied.

Here we present an example of a hollow pentagon moving under kinematic control and surrounded internally and externally with fluid. It is centered on the origin and has vertices at $(L\cos(\frac{2\pi k}{5}+\frac{\pi}{10}),L\sin(\frac{2\pi k}{5}+\frac{\pi}{10}))$ for $k\in\mathbb{Z}$, with outer radius $L=\SI{0.25}{\cm}$ and inner radius $L=\SI{0.2}{\cm}$. The fluid has viscosity $\mu_f=\SI{0.01}{\pascal\second}$, and density $\rho_f=\SI{1.0}{\gram/\cm^3}$, and the pentagon has density $\rho_r=\SI{1.0}{\gram/\cm^3}$. The resolution is $200\times 200$, the simulation domain is $[-0.5,0.5]\times[-0.5,0.5]\,\si{\cm^2}$ and no-slip and no penetration boundary conditions are used. The fluid and rotor are initially stationary. A rotational motion is prescribed for the rotor with angular velocity $\lambda=-2\sin(10\pi t)\text{~rad/s}$, and the center of the rotor is fixed at the origin so that $\vu_T=(0,0)\,\si{cm/s}$. Snapshots of vorticity $\omega$ are shown in Fig.~\ref{fig:pentagon}. The pentagon stays rigid during the simulation, as indicated by a spatially constant vorticity field inside the object. The sign of the vorticity also indicates the rotational direction of the pentagon. When $\omega$ is negative, it rotates counterclockwise in time intervals $[0,0.1]\,\si{s}$, and when $\omega$ is positive, it rotates clockwise in intervals $t=[0.1,0.2]\,\si{s}$.

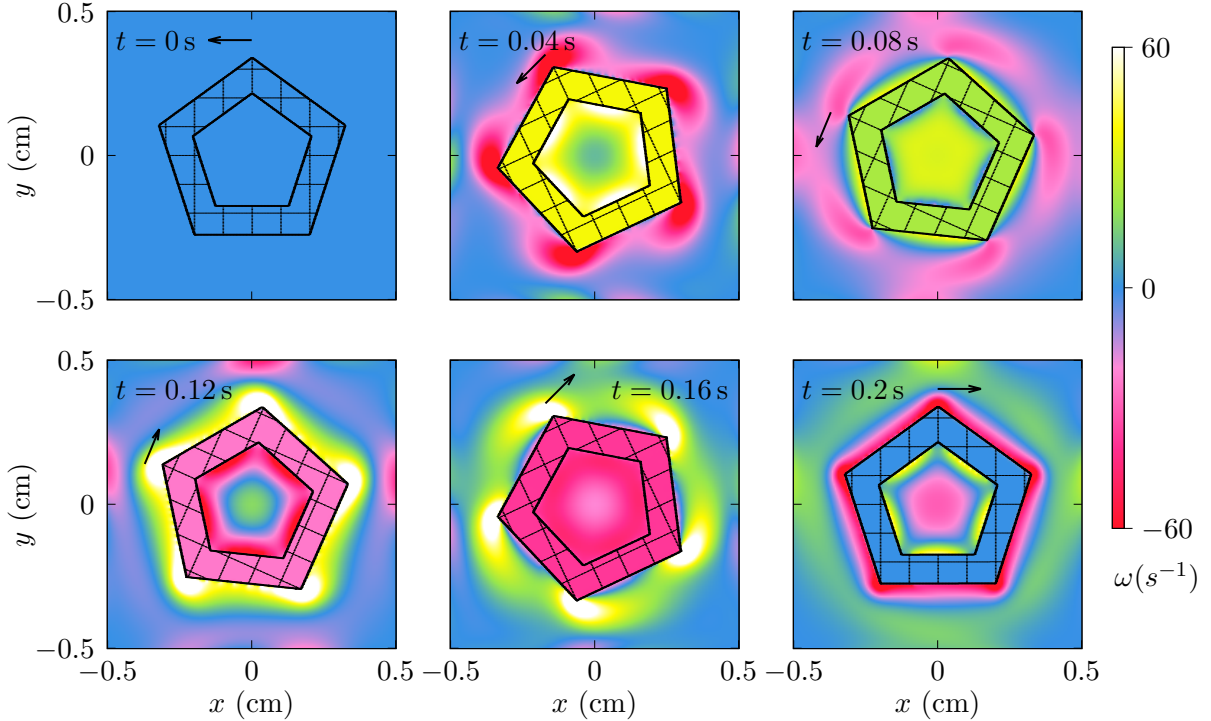
\Eoneplots{pentagon}{\label{fig:pentagon}Snapshots of vorticity $\omega$ in a simulation of a rigid hollow pentagon being spun with a prescribed motion $\vu_T=(0,0) \,\si{\cm/s},\lambda=-2\sin(\pi t) \text{~rad/s}$. Other simulation parameters are $\rho_f=\rho_r=\SI{1.0}{\gram/\cm^3}$ and $\mu_f=\SI{0.01}{\pascal\second}$. The arrows indicate the direction of rotation at each snapshot.}

With a small modification, kinematic boundary conditions can also be applied to soft structures using our projection approach. In previous work~\cite{rycroft2018reference}, a kinematic condition on a small part of a soft object was implemented using a penalty method, which introduces another spring constant into the system, whose magnitude is bounded by the time step to ensure a stability requirement. What we propose now is an alternative approach that implicitly enforces such constraints while allowing the rest of the soft solid to deform passively. 

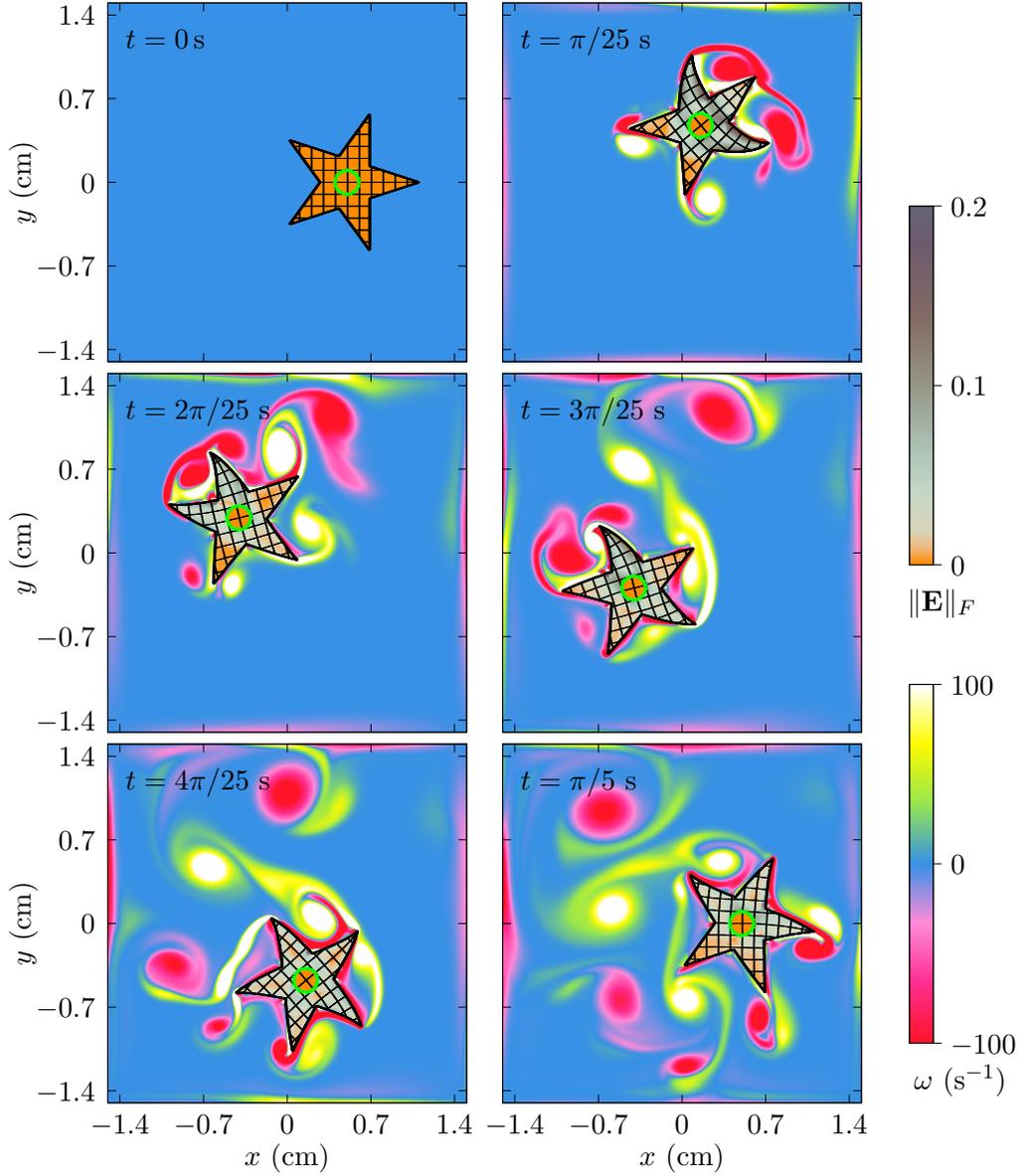
\Eoneplots{five-star}{\label{fig:five-star}Snapshots of vorticity $\omega$ and the Frobenius norm of the Hencky strain $\|\mathbf{E}\|_F$ in a simulation of a soft five-pointed rotor being rotated and translated with a prescribed motion. The vorticity field is plotted in the fluid domain and the norm of the strain tensor is plotted in the structure domain. The thick black line marks the fluid--structure interface. The dashed lines inside the structure are the contours of the components of the reference map. The star shape has a outer radius of $\SI{0.4}{\cm}$ and shear modulus $G=\SI{300}{\pascal}$. A smaller inner circular area with radius $\SI{0.1}{\cm}$ is set to be rigid where the extra body forces are computed to enforce an prescribed motion: $u_T=-0.5\sin(t)\,\si{\cm/s},v_T=0.5\cos(t)\,\si{\cm/s},\lambda=\sin(t)\,\si{rad/s}$. The soft--rigid interface is shown with green dots. Other simulation parameters used here are $\rho_f=\rho_s=\rho_r=\SI{1.0}{\gram/\cm^3}$, and $\mu_f=\SI{0.01}{\pascal\second}$. }

In figure \ref{fig:five-star}, we present an example of this capability using a soft five-pointed star in a fluid box $[-1.5,1.5]\times[-1.5,1.5]\,\si{\cm^2}$, with density $\rho_f=\SI{1.0}{\gram/\cm^3}$ and viscosity $\mu_f=\SI{0.01}{\pascal\second}$. Initially, the five-pointed star is centered at $(0.5,0)\,\si{cm}$ and has a vertex at $(L\cos\frac{2\pi k}{5},L\sin\frac{2\pi}{5})$ with $L=\SI{0.4}{\cm}$, density $\rho_s=\SI{1.0}{\gram/\cm^3}$, and shear modulus $G=\SI{300}{\pascal}$. An inner circular rigid domain with radius $r=\SI{0.1}{\cm}$ is defined around the center of the star shape, with the same solid density $\rho_r=\SI{1.0}{\gram/\cm^3}$, and both a translation velocity $(u_T,v_T) = (0.5\sin (10t),0.5\cos(10t))\,\si{\cm/s}$ and an angular velocity $\lambda =\sin(10t)\,\si{rad/s}$ are prescribed in the rigid domain to force the star to move. The vorticity field $\omega$ is shown in the fluid domain, and the Frobenius norm of the Hencky strain $\|\mathbf{E}\|_F$ is plotted in the structure to demonstrate the deformation of the structure during the simulation. The rigid domain has zero strain tensor during the simulation as expected.

\section{Conclusion}
In this work, we have presented an Eulerian framework combining the reference map technique and a projection method to study a variety of FSI problems for mixed soft and rigid objects. It is an extension to our previous soft FSI solver and takes full advantage of the features that were already developed in the previous work. It allows a much larger time constraint for simulating rigid objects, which is particularly useful for stiff problems where rigid objects are mixed with very soft objects. The incompressibility and rigidity constraints are implemented as spatially linear constraints on an Eulerian grid, which is simpler than how these constraints would appear if handled in Lagrangian frame.  We are able to maintain an overall second-order accuracy of solutions in $L_2$ norm, although the accuracy drops directly adjacent to the fluid-structure interface. We expect this can be improved with an adaptive mesh technique near the fluid-solid interface. 

Our simulation results are shown to match other numerical and analytical results for a single circular object sedimenting across the low and moderate Reynolds number regime, and over a variation of fluid/solid density ratios. We also demonstrated that the model can be extended to solve soft structures with rigid inner components submerged in fluids, which has potential for applications in materials science and biomechanics. Additional capacity for simulating contact between multiple rigid and soft objects is also demonstrated. It accurately captures the simultaneous and complex interactions between fluids, soft structures, and rigid structures, yet maintains the computational simplicity of working with a single fixed Eulerian grid. With a similar methodology, we also presented how to control the kinematics of subdomains of soft structures in a fully coupled FSI model via the projection framework, which only requires a few more equations in the projection step.

There are several future directions to explore for this framework. With the ability to deal with multi-phase interactions between fluids, rigid and soft objects, and with the flexibility of controlling the kinematics of structures, we foresee the application of this simulation model to assist experimental studies. There are also opportunities to model solids beyond elasticity, which can be done by combining new state variables and constitutive relations. This model naturally has the potential to generalize to three dimensions, in which case the symmetric and traceless rigid stress will have five independent components. The rigid stress plus the pressure will have six independent components to solve at the projection step. The rigid constraint and the incompressible constraint again provide six equations with a common Laplacian operator applying on the pressure term. A challenge will be to formulate the linear matrix at the projection step to maintain an SPD system for efficient computation, and this may be the subject of future work.
\section*{Data Availability Statement}
The data that support the findings of this study are available from the corresponding author upon reasonable request.
\section*{Acknowledgements}
X.W.~and K.K.~were partially supported by ARO grants  W911NF-19-1-0431 and W911NF-16-1-0440. C.H.R.~was partially supported by the Applied Mathematics Program of the U.S. DOE Office of Science Advanced Scientific Computing Research under contract number DE-AC02-05CH11231. 

\appendix
\section{Linear system in the projection step}\label{section:spd}
For a given pressure element $\psi$ and rigid stress element $\gamma$, the weak formulation of Eqs.~\eqref{eq:incomp}, \eqref{eq:dev1}, \& \eqref{eq:dev2} in the projection step are
\bqa
-\int_\Omega\frac{\Delta t}{\rho(\phi^{n+\frac{1}{2}})}\nabla p^{n+1}\cdot\nabla\psi \,dx\,dy+\int_{\Omega'_r}\frac{\Delta t}{\rho(\phi^{n+\frac{1}{2}})}\left(\left[\begin{array}{c}\theta_x\\-\theta_y\end{array}\right]+\left[\begin{array}{c}\tau_y\\ \tau_x\end{array}\right]\right)^{n+1}\cdot\nabla\psi\ dx\,dy \nonumber\\
= \int_\Omega-(\nabla\cdot\vu^*)\psi\ dx\,dy,\label{eq:appweak1}\\
\int_{\Omega'_r}\frac{\Delta t}{\rho(\phi^{n+\frac{1}{2}})}\left(\left[\begin{array}{c}p_x\\-p_y\end{array}\right]-\nabla\theta\right)^{n+1} \cdot\nabla\gamma\ dx\,dy= \int_{\Omega'_r}(u^*_x-v^*_y)\gamma\ dx\,dy,\label{eq:appweak2}\\
\int_{\Omega'_r}\frac{\Delta t}{\rho(\phi^{n+\frac{1}{2}})}\left(\left[\begin{array}{c}p_y\\p_x\end{array}\right]-\nabla\tau\right)^{n+1}\cdot\nabla\gamma\ dx\,dy = \int_{\Omega'_r}(u^*_y+v^*_x)\gamma\ dx\,dy,\label{eq:appweak3}
\eqa
Consider a particular bilinear element function $\psi$ located at pressure point $p_{i,j}$ and $\gamma$ located at rigid stress point $\theta_{i,j}$ and $\tau_{i,j}$ where the point $(x_i,y_j)$ is in the rigid domain. In the simpler case when $\rho$ is constant (so that the fluid density equals the structure density), the first term in Eq.~\eqref{eq:appweak1} is essentially a Laplacian operator on $p$, defined as
\bq
L_\Omega (p_{i,j})\triangleq\lambda_ap_{i,j}+\lambda_b(p_{i-1,j}+p_{i+1,j})+\lambda_c(p_{i,j-1}+p_{i,j+1})+\lambda_d\sum\limits_{\substack{k=\pm 1\\l=\pm 1}}p_{i+k,j+l}
\eq
where
\bq
\lambda_a=\displaystyle\frac{4(\Delta x^2+\Delta y^2)}{3\Delta x\Delta y},\ \lambda_b=\displaystyle\frac{\Delta x^2-2\Delta y^2}{3\Delta x\Delta y},\
\lambda_c=\displaystyle\frac{-2\Delta x^2+\Delta y^2}{3\Delta x\Delta y},\
\lambda_d=\displaystyle\frac{-\Delta x^2-\Delta y^2}{6\Delta x\Delta y}.
\label{eq:lambdas}
\eq
The second and third terms in Eq.~\eqref{eq:appweak1} on $\theta$ and $\tau$ are
\bqa
W_{\Omega'_r} (\theta_{i,j})\triangleq\beta_a\theta_{i,j}+\beta_b(\theta_{i-1,j}+\theta_{i+1,j})+\beta_c(\theta_{i,j-1}+\theta_{i,j+1})+\beta_d\sum\limits_{\substack{k=\pm 1\\l=\pm 1}}\theta_{i+k,j+l},\\
R_{\Omega'_r}(\tau_{ij})\triangleq\frac{1}{2}(\tau_{i+1,j+1}-\tau_{i+1,j-1}+\tau_{i-1,j+1}-\tau_{i-1,j-1}),
\eqa
where
\bq
\beta_a=\displaystyle\frac{4(-\Delta x^2+\Delta y^2)}{3\Delta x\Delta y},\ \beta_b=\displaystyle\frac{\Delta x^2+2\Delta y^2}{3\Delta x\Delta y},\
\beta_c=\displaystyle\frac{-2\Delta x^2-\Delta y^2}{3\Delta x\Delta y},\
\beta_d=\displaystyle\frac{\Delta x^2-\Delta y^2}{6\Delta x\Delta y}.
\eq
The discretized equation can be written as
\bq
\left(\begin{array}{ccc}
L_\Omega &W_{\Omega'_r} & R_{\Omega'_r} \\
W_{\Omega'_r}^\trans &L_{\Omega'_r,0} &0\\
R_{\Omega'_r}^\trans &0&L_{\Omega'_r,0}
\end{array}\right)\left(\begin{array}{c}p\\ \theta\\ \tau\end{array}\right)=\text{RHS}\label{eq:stiffmatrix}
\eq
where $L_{\Omega'_r,0}$ is the Poisson operator confined on $\Omega'_r$ with zero boundary condition. The vector on the right hand side (RHS) of Eq.~\eqref{eq:stiffmatrix} is
\bqa
\left(\begin{array}{c}-\Delta x(u^*_{i+1,j+1}+u^*_{i+1,j}-u^*_{i,j+1}-u^*_{i,j})-\Delta y(v^*_{i+1,j+1}-v^*_{i+1,j}+v^*_{i,j+1}-v^*_{i,j})\\
\Delta x(u^*_{i+1,j+1}+u^*_{i+1,j}-u^*_{i,j+1}-u^*_{i,j})-\Delta y(v^*_{i+1,j+1}-v^*_{i+1,j}+v^*_{i,j+1}-v^*_{i,j})\\
\Delta y(u^*_{i+1,j+1}-u^*_{i+1,j}+u^*_{i,j+1}-u^*_{i,j})+\Delta x(v^*_{i+1,j+1}+v^*_{i+1,j}-v^*_{i,j+1}-v^*_{i,j})
\end{array}
\right).
\eqa
This linear system is symmetric. Now we show it is also positive definite.
We express the variables as $p=\sum_{i=1}^{K}p_i\psi_i$ where $K=(M+1)\times (N+1)$ for non-periodic boundary conditions, and $\theta=\sum_{j=1}^T\theta_j\gamma_j$, $\tau=\sum_{i=1}^T\tau_i\gamma_i$, where $T$ is the number of the rigid points, and $T< K$ as the rigid solid is fully immersed in the fluid domain. For all $\vq=(p,\theta,\tau)^\trans$ we have
\bqa
&&\vq^\trans\left(\begin{array}{ccc}
L_\Omega &W_{\Omega'_r} & R_{\Omega'_r} \\
W_{\Omega'_r}^\trans &L_{\Omega'_r,0} &0\\
R_{\Omega'_r}^\trans &0&L_{\Omega'_r,0}
\end{array}\right)\vq=\begin{array}{c}p L_\Omega(p)+\theta L_{\Omega'_r,0}(\theta)+\tau L_{\Omega'_r,0}(\tau)\\
+p W_{\Omega'_r}(\theta)+\theta W_{\Omega'_r}^\trans(p)+p R_{\Omega'_r}(\tau)+\tau R_{\Omega'_r}^\trans(p)
\end{array} \\
&&=\displaystyle\left(\sum\limits_{i,j=1}^K\int_{V_{ij}^h}p_ip_j\nabla\phi_i\cdot\nabla\phi_j \,dx\,dy\right) +\left(\sum\limits_{\substack{i,j=1\\ \vx_i,\vx_j\in\Omega'_r}}^{T} \int_{V_{ij}^h}(\theta_i\theta_j+\tau_i\tau_j)\nabla\gamma_i\cdot\nabla\gamma_j\,dx\,dy\right)\nonumber\\
&&+\sum\limits_{\substack{i,j=1\\ \vx_i,\vx_j\in\Omega'_r}}^{T} \int_{V_{ij}^h}p_i\left(\theta_j\left[\begin{array}{c}-\gamma_{jx}\\ \gamma_{jy}\end{array}\right]+\tau_j\left[\begin{array}{c}-\gamma_{jy}\\ -\gamma_{jx}\end{array}\right]\right)\cdot\nabla\phi_i \,dx\,dy\nonumber \\
&&+\sum\limits_{\substack{i,j=1\\ \vx_i,\vx_j\in\Omega'_r}}^{T} \int_{V_{ij}^h}p_j\left(\theta_i\left[\begin{array}{c}-\phi_{jx}\\ \phi_{jy}\end{array}\right]+\tau_i\left[\begin{array}{c}-\phi_{jy}\\ -\phi_{jx}\end{array}\right]\right)\cdot\nabla\gamma_i \,dx\,dy \nonumber \\
&&=\sum\limits_{\substack{i=1\\ \vx_i\notin\Omega'_r}}^K\int_{V^h_{i}}p_i^2\|\nabla\phi_i\|^2dxdy +\sum\limits_{\substack{i=1\\ \vx_i\in\Omega'_r}}^T\int_{V_{i}^h}(p_i\phi_{ix}-\theta_i\gamma_{ix}-\tau_i\gamma_{iy})^2+(p_i\phi_{iy}+\theta_i\gamma_{iy}-\tau_i\gamma_{ix})^2\,dx\,dy \nonumber\\
&&\geq 0.
\eqa
Here $V_{ij}^h$ is the discretized space where two basis functions (\textit{e.g.}, $\phi_i$ and $\phi_j$ or $\gamma_i$ and $\gamma_j$) overlap. The equals sign holds when each integral is zero. This requires the pressure field to be all zero outside the rigid domain, and all the terms in the second summation to be zero. For bilinear basis functions and other higher order basis functions, this implies that
\bqa
p_i-\theta_i=0,&\ \tau_i=0,&\ \forall i=1,\ldots,T\\
p_i+\theta_i=0,&\ \tau_i=0,&\ \forall i=1,\ldots,T,
\eqa
which indicates that equality holds only when $\vq=\mathbf{0}$. Hence the linear system is SPD.
\section{Tests of convergence}\label{sec:conv-test}

To study the accuracy of the numerical method, we performed a convergence test where a rigid circle was released in a fluid box of $[-0.5,0.5]\times[-0.5,0.5]\si{\cm^2}$. The circle was centered at $(0.1,0.1)$ at the beginning with a radius of $r=0.12$, and was released with an initial velocity of $(-5,-5)\si{\cm/\second^2}$ without body forces. In the simulation, we chose $G_{\text{col}}=10\si{\pascal}$, $\rho_r=\rho_f=1.0\si{\gram/\cm^3}$, and $\mu_f=1\times 10^{-3}\si{\pascal\second}$. We are mostly interested in the spatial accuracy of our method since the temporal accuracy of our method inherits the properties of the projection method. A fixed time step $\Delta t=2\times 10^{-5}$ was chosen so that it meets the time restriction requirements at the finest resolution.

Since the problem has no analytical solution, we performed reference simulations using a $720\times 720$ grid. We then follow the standard procedure for numerical methods by running a set of coarser simulations using $N\times N$ grids where $N\in\{90,120,180,240,360\}$ to compare against the reference results. We note that each $N$ divides evenly by 720, therefore the grid corners of these coarser simulations align with the reference simulation.  In our model, the blur zone and the extra viscosity both scale with the grid size, which is consistent with the convergence study in other numerical methods \cite{leveque2002finite}. However, we noted that under this set up the discretized problem solved in coarser and reference grids are not exactly the same, and we expect this will affect the convergence rate obtained from this study, particularly near the fluid-solid interface. Therefore, we also consider another study where the extra viscosity and blur width are both fixed. In particularly, these two values are chosen based on the coarsest grid ($90\times 90$).

We consider normalized error measures with respect to $L_q$ norms where
\begin{equation}
    E_q^{\mathbf{v}}=\left(\frac{1}{A}\int_\Omega \|\mathbf{v}_{\text{ref}}-\mathbf{v}\|_2^q\mathbf{dx}\right)^{1/q}
\end{equation}
where $A=1$ is the area of the domain, and the `ref' subscript refers to the reference velocity field. The integral is calculated using a direct sum over the field values. The velocity field is cell-centered, so the reference simulation result is interpolated into those points using a bilinear interpolation, which results in a $O(\Delta x^3)$ error. 

The results of the convergence study are shown in figure \ref{fig:conv-test}, where the $L_2$-norm error is presented in panel (a) and $L_\infty$-norm error is presented in panel (b). `Constant blur' refers to the case where the extra viscosity and the blur width were both fixed. Dashed lines with slope $p=2$ and $p=1.5$ were also plotted to illustrated the convergence rate of the methods, respectively in panel (a) and (b). We noted that for both studies, the convergence rate for $L_\infty$ norm is lower as the error was concentrated near the fluid-solid interface. The `standard' study has a convergence rate $\approx 1.5$ with respect to the $L_2$ norm, while the `constant blur' case achieved a second order convergence rate. 
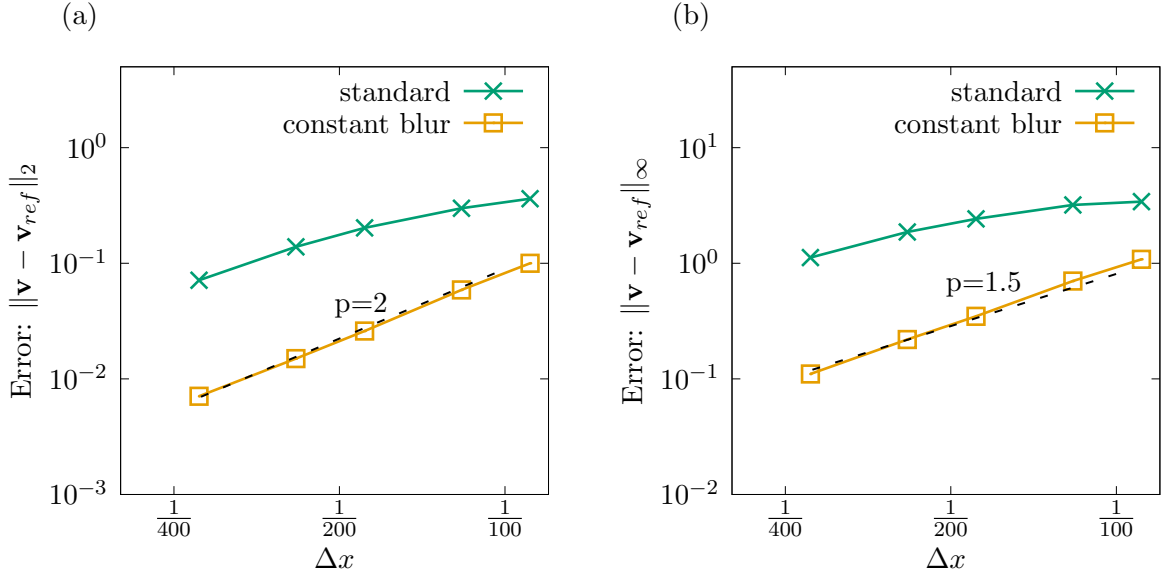
\Eoneplots{conv-test}{\label{fig:conv-test} Plots showing the convergence rate of the solutions for different grid sizes $\Delta x$ for the two convergence tests. `Constant blur' refers to the case where the extra viscosity and the blur width were both fixed. The errors measured in $L_2$ and $L_\infty$ norms are shown in panel (a) and (b). Dashed lines with slope $p=2$ and $p=1.5$ were also plotted to illustrated the convergence rate of the methods, respectively in panel (a) and (b). }

In our previous work \cite{rycroft2018reference}, the effect of the blur zone was discussed thoroughly where we showed the size of the blur zone is a trade-off between the additional noise for small $\epsilon$, and excessive blurring for large $\epsilon$. We refer the readers for more details in that work \cite{rycroft2018reference}.

\bibliography{EulerianMethod}
\end{document}